\documentclass[11pt]{article}
\usepackage[margin=1 in]{geometry}
\usepackage{setspace}
\usepackage{amsthm,amssymb,amsmath,euscript,mathrsfs,dsfont}
\usepackage{times}
\usepackage{bm,bbm}
\usepackage{cases}
\usepackage[round]{natbib}
\usepackage{booktabs}  
\usepackage{threeparttable}
\usepackage{makecell}
\usepackage[plain,noend]{algorithm2e}
\usepackage{euscript}
\usepackage{verbatim}
\usepackage{graphicx}
\usepackage{enumitem}
\usepackage{color}
\usepackage{verbatim} 
\usepackage{rotating}
\usepackage{multirow}
\usepackage[width=\textwidth]{caption}
\usepackage{hyperref}
\hypersetup{colorlinks=true,linkcolor=blue,urlcolor=blue,citecolor=blue}
\usepackage[title]{appendix}
\usepackage[dvipsnames]{xcolor} %
\usepackage{authblk}
\usepackage{comment}
\usepackage{tikz}
\newtheorem{proposition}{Proposition}
\newtheorem{lemma}{Lemma}
\newtheorem{theorem}{Theorem}
\newtheorem{definition}{Definition}

\newtheorem{remark}{Remark}
\newtheorem{algo}{Algorithm}

\newcommand{\RN}[1]{\textup{\uppercase\expandafter{\romannumeral#1}}}

\newcommand{\argmin}{\operatornamewithlimits{argmin}}

\newcommand{\blind}{1} 

\begin{document} 

\doublespace

\date{}

\if1\blind
{
  \title{\bf Multiple Testing of Local Extrema for Detection of Structural Breaks in Piecewise Linear Models}
  \author{Zhibing He, Dan Cheng\footnote{Corresponding author. Email: chengdan@asu.edu. This paper has been accepted by the Statistica Sinica.}\;  and Yunpeng Zhao \\
    Yale University, Arizona State University and  Colorado State University} 
  \maketitle
} \fi

\if0\blind
{
  \bigskip
  \bigskip
  \bigskip
  \begin{center}
    {\LARGE\bf Multiple Testing of Local Extrema for Detection of Structural Breaks in Piecewise Linear Models}
\end{center}
  \medskip
} \fi

\bigskip

\begin{abstract}
In this paper, we propose a novel method for detecting the number and locations of structural breaks or change points in piecewise linear models. By transforming change point detection into identifying local extrema through kernel smoothing and differentiation, we compute p-values for the local extrema and apply the Benjamini-Hochberg (BH) procedure to identify significant local extrema as the detected change points. Our method effectively distinguishes between two types of change points: continuous breaks (Type \RN{1}) and jumps (Type \RN{2}). We study three scenarios of piecewise linear signals: pure Type \RN{1}, pure Type \RN{2} and a mixture of both. The results demonstrate that our proposed method ensures asymptotic control of the False Discovery Rate (FDR) and power consistency as sequence length, slope changes, and jump size increase. Furthermore, compared to traditional recursive segmentation methods, our approach requires only one multiple-testing step for candidate local extrema, thereby achieving the smallest computational complexity proportionate to the data sequence length. Additionally, numerical studies illustrate that our method maintains FDR control and power consistency, even in non-asymptotic situations with moderate slope changes or jumps.
We have implemented our method in the R package \href{https://cran.r-project.org/web/packages/dSTEM}{``dSTEM''}.
\end{abstract}
\noindent%
\vfill

\newpage


\section{Introduction}
\label{sec:intro}

\begin{figure}[!ht]
	\centering
    \includegraphics[width=0.80\textwidth]{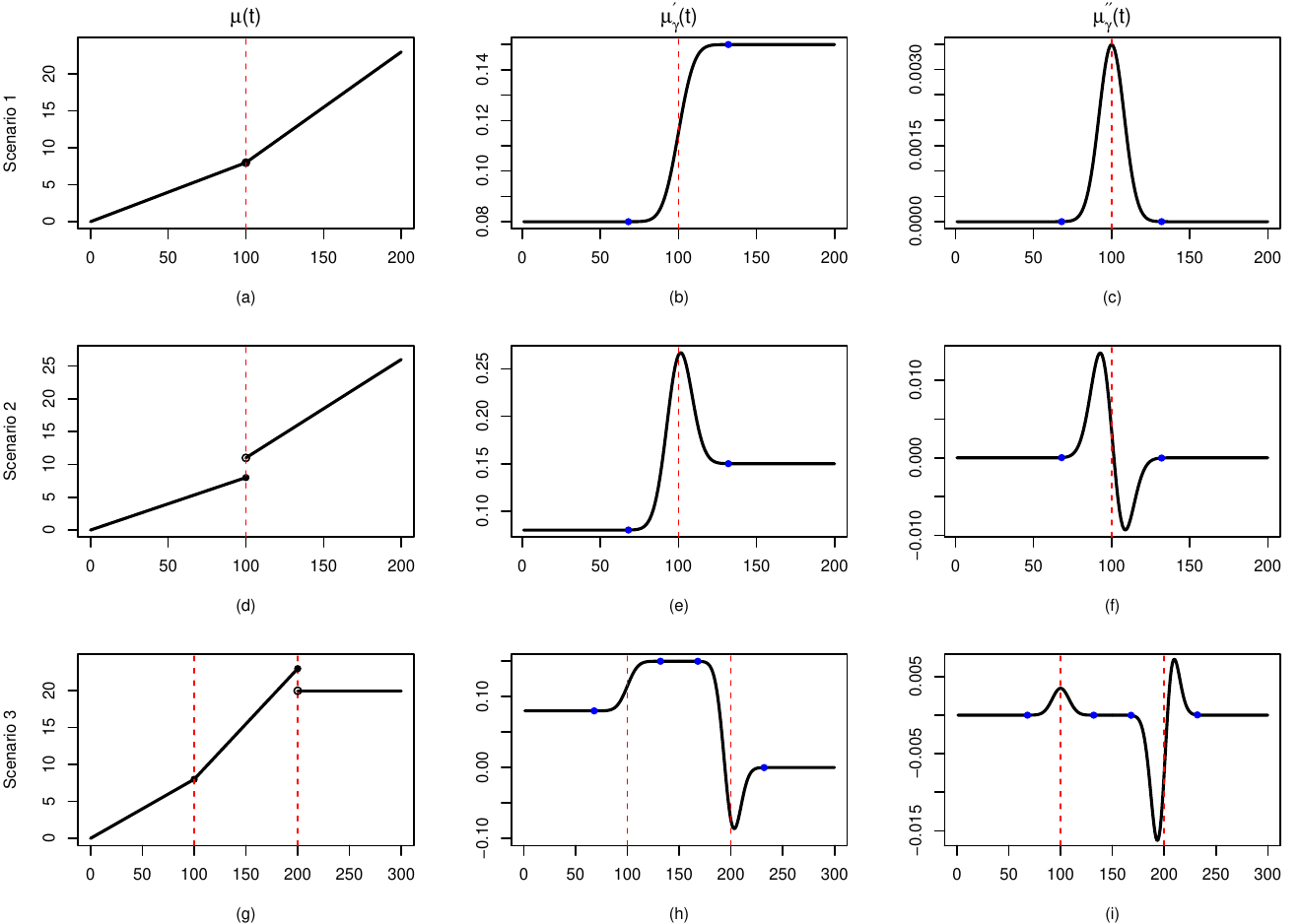}
	\caption{\footnotesize {Illustration of change point detection. 
	A change point in the piecewise linear signal $\mu(t)$ (left panel) becomes a local extremum in either the first derivative of the smoothed signal $\mu'_{\gamma}(t)$ (middle panel) or in the second derivative of the smoothed signal $\mu''_{\gamma}(t)$ (right panel).
	The red dashed lines indicate the locations of true change points, and
	the blue points represent the regions where the signal is smoothed using a Gaussian kernel.
	The top row shows a Type \RN{1} change point $v_j$ in $\mu(t)$ becomes a local extremum in $\mu''_{\gamma}(t)$ precisely at $t = v_j$.
	The middle row illustrates a Type \RN{2} change point $v_j$ in $\mu'_{\gamma}(t)$ becomes a local extremum in $\mu''_{\gamma}(t)$ around $v_j$.
	The bottom row shows that only Type \RN{2} change points can generate local extrema in $\mu'_{\gamma}(t)$, whereas both Type \RN{1} and Type \RN{2} change points can generate local extrema in $\mu''_{\gamma}(t)$.}}
	\label{fig:illu1}
\end{figure}

In this paper, we consider a canonical univariate statistical model: 
\begin{equation}
\label{eq:signal+noise}
y(t) = \mu(t) + z(t), \quad t \in \mathbb{R},
\end{equation}
where $z(t)$ is random noise and $\mu(t)$ is a piecewise linear signal of the form
$
\mu(t) = c_j + k_j t, \quad t\in (v_{j-1}, v_j], 
$
where $c_j, k_j\in \mathbb{R}$, $j=1, 2, \dots$ and $- \infty =v_0<v_1<v_2<\cdots$. 
Assume that the structures of $\mu(t)$ are different in the neighboring $v_j$, i.e., $(c_j, k_j) \neq (c_{j+1}, k_{j+1})$, resulting in a continuous break or jump at $v_j$ (see Figure \ref{fig:illu1}). This $v_j$ is called a change point or structural break.

The detection of structural breaks or change points plays a pivotal role in various fields, including statistics, econometrics, genomics, climatology, and medical imaging. These detection methods are broadly applied to different domains based on different types of signals. For example, in monitoring medical conditions \citep{khan2020overview}, the presence of piecewise constant signals with jumps can indicate significant changes in patient health conditions. Accurate detection of these changes is essential for aiding in diagnosis, treatment planning, and object recognition tasks. On the other hand, piecewise linear signals with continuous breaks are commonly encountered in climate change research \citep{tebaldi2008towards}. These changes may represent gradual shifts in temperature patterns or human behavior. Furthermore, in financial markets, piecewise linear signals with non-continuous breaks may indicate sudden shifts in stock prices or market dynamics \citep{chang2008integrating}. Detecting these changes is crucial for making informed investment decisions.

To differentiate between the various ways of linear structural changes, we define the following two types of change points.
\begin{definition}\label{def:break}
A point $v_j$ is called a Type \RN{1} change point (structural break) if $c_j + k_j v_j = c_{j+1} + k_{j+1}v_j$ and $k_j \neq k_{j+1}$, and a point $v_j$ is denoted as a Type \RN{2} change point if $c_j + k_jv_j \neq c_{j+1} + k_{j+1}v_j$ for $j \geq 1$.	
\end{definition}

At Type \RN{1} change points $v_j$, the signals remain continuous, but the slopes experience a sudden change at $v_j$. 
Type \RN{2} change points correspond to jumps in the signal, where there is a discontinuity between adjacent segments. Note that a special case of Type \RN{2} change points is that $\mu(t)$ is piecewise constant, i.e., $k_j \equiv 0$ and $c_j \neq c_{j+1}$ for $j \geq 1$. 
In this paper, we specifically focus on three scenarios of the signal $\mu(t)$, each representing different characteristics of the change points. The scenarios are described as follows:

\noindent \textbf{Scenario 1: Pure Type \RN{1} Change Points.} The signal $\mu(t)$ consists of continuous segments with slope changes at specific points $v_j$.  Figure \ref{fig:illu1} (a) illustrates this scenario, where the signal exhibits continuous behavior with distinct slopes at the change points.\\
\noindent \textbf{Scenario 2: Pure Type \RN{2} Change Points.} The signal $\mu(t)$ contains only Type \RN{2} change points (jumps). Figure \ref{fig:illu1} (d) illustrates this scenario, where the signal exhibits sudden changes or discontinuities at the change points.\\
\noindent \textbf{Scenario 3: Mixture of Type \RN{1} and Type \RN{2} Change Points.} The signal $\mu(t)$ combines both continuous breaks (Type \RN{1}) and jumps (Type \RN{2}). Figure \ref{fig:illu1} (g) illustrates this scenario, which allows a more general representation of signals with mixed characteristics.

Our primary objective is to detect both the number of change points and their locations simultaneously. 
To address this problem, we propose a novel approach that can handle an unknown number of multiple structural breaks occurring at unknown positions within the signal. 
Our proposed method for change point detection is illustrated and evaluated in the three aforementioned scenarios (pure Type \RN{1} change points, pure Type \RN{2} change points, and a mixture of Type \RN{1} and Type \RN{2} change points).  Figure \ref{fig:illu1} shows the application of our method to each scenario, showcasing the mechanism behind our change point detection method.

The key idea behind our approach is based on the observation that a change point in the signal $\mu(t)$ will manifest as a local extremum (local maximum or local minimum) in either the first or second derivative of the smoothed signal. By kernel smoothing and differentiating the data sequence, we transform the problem of change point detection into identifying local extrema. Specifically,
a Type \RN{1} change point in $\mu(t)$ becomes a local extremum in the second derivative $\mu''_{\gamma}(t)$ (see Figure \ref{fig:illu1} (a) and (c)).  A Type \RN{2} change point in $\mu(t)$ becomes a local extremum in the first derivative $\mu'_{\gamma}(t)$ (see Figure \ref{fig:illu1} (d) and (e)).
Therefore, in Scenario 1 and Scenario 2, the detection of change points relies on identifying local extrema in the second and first derivatives of the smoothed signal, respectively. 
However, in Scenario 3, a different approach is required due to the unique characteristics of each type.
Type \RN{1} change points do not generate local extrema in the first derivative $\mu'_{\gamma}(t)$ (see Figure \ref{fig:illu1} (b)). Therefore, in Scenario 3, the detection process begins by identifying all Type \RN{2} change points as local extrema in $\mu'_{\gamma}(t)$.
On the other hand, local extrema in the second derivative $\mu''_{\gamma}(t)$ can be generated by both Type \RN{1} and Type \RN{2} change points (see Figure \ref{fig:illu1} (i)). To detect Type \RN{1} change points in Scenario 3, one needs to remove the local extrema in $\mu''_{\gamma}(t)$ generated by Type \RN{2} change points.
By combining these approaches, our proposed method can distinguish between Type \RN{1} and Type \RN{2} change points. 

The detection of local extrema can be formulated as a peak detection problem, focusing on the local extrema of the first and second derivatives. In the literature, there have been notable contributions that address this issue. One such notable approach is the Smoothing and TEsting of Maxima (STEM) algorithm \citep{schwartzman2011multiple}.
The STEM algorithm aims to identify local maxima of the derivative as candidate peaks, which correspond to potential change points in the signal. By smoothing the signal with the kernel and applying multiple tests, the algorithm distinguishes between local extrema generated by true change points and those arising from random noise.
Moreover, \citet{cheng2020multiple} introduced the differential STEM (dSTEM) method, specifically designed to detect change points in data sequences modeled as a piecewise constant signal plus noise. The dSTEM method leverages the principles of the STEM algorithm but adapts it to the piecewise constant signal model. This modification allows for more effective change point detection in scenarios where the signal exhibits piecewise constant behavior.

The literature on change point detection contains a large amount of statistical inference research, but most of it is specifically designed for the case of a single change point with an unknown location. For example, in the case of a single change point with an unknown location, \citet{andrews1993tests} proposed comprehensive treatments and testing methods for structural change. \citet{perron1989great} utilized unit root tests to detect a one-time change in the level or slope of the trend function in univariate time series. \citet{bai1994least} introduced the least squares method for estimating an unknown shift point (change point) in a piecewise constant model.
In recent years, there has been extensive interest in multiple change point detection, particularly in the context of multiple testing-based methods. In the piecewise constant signal model, \citet{yao1989least} studied least squares estimators for the locations and levels of the step function under both known and unknown numbers of jumps. \citet{lavielle2005using} developed a penalized least squares method to estimate the number of change points and their locations. Binary segmentation (BS) \citep{vostrikova1981detecting} is another popular approach, \citet{hyun2021post} outlined similar post-selection tests for change point detection using Wild Binary Segmentation  (WSB) \citep{fryzlewicz2014wild} and circular Binary Segmentation (CBS) \citep{olshen2004circular}.
SMUCE \citep{frick2014multiscale} estimates the number of change points as the minimum among all candidate fits, where the empirical residuals pass a certain multi-scale test at a given significance level. \citet{li2016fdr} proposed an estimator that controls the False Discovery Rate (FDR) while allowing for a generous definition of a true discovery. FDR control was also demonstrated for the SaRa estimator \citep{hao2013multiple} and the dSTEM estimator \citep{cheng2020multiple} for multiple change point locations.
For the continuous (Type I) change point, the underlying model corresponds to the classical bent–line (segmented) regression framework, where two linear segments meet at an unknown knot. A rich literature exists on this problem, including likelihood–based and grid–search approaches \citep{lerman1980fitting}, as well as more recent computationally efficient estimation methods such as the iterative procedure of \citet{muggeo2003estimating}.
Additionally, \citet{Bai1998} provided asymptotic distribution results regarding the distance between estimated change points and their true locations, assuming a known number of change points.
Furthermore, methods such as narrowest-over-threshold (NOT) \citep{baranowski2019narrowest} and narrowest significance pursuit (NSP) \citep{fryzlewicz2020narrowest} have been proposed to detect change points by focusing on localized regions that contain suspected features.
These works represent a diverse range of approaches for change point detection, providing valuable insights and methodologies for addressing both single and multiple change point scenarios in various signal models.

In this paper, we propose a modified Smoothing and TEsting of Maxima (mSTEM) algorithm for change point detection in piecewise linear models.
Our proposed approach is unique compared to the existing literature in the following ways:
1. 
Our method allows for the simultaneous estimation of both the number and locations of change points; 
2. 
Our method can effectively distinguish between Type \RN{1} and Type \RN{2} change points;
3. Our proposed method achieves significantly lower computational complexity compared to traditional change point detection methods. 
By testing only the candidate peaks generated by true change points or random noise once, the method reduces the computation to the number of candidate peaks, which is proportional to the length of the data sequence;
4. 
We derive a general and straightforward form of the peak height distribution for derivatives of Gaussian processes, which depends on only one parameter. This generalizes and extends the distributions used in \citet{schwartzman2011multiple};
5. 
Our method takes into account the presence of correlated noise by assuming that the noise follows a Gaussian process.

This paper is organized as follows. Section \ref{sec:mSTEM} introduces the framework of our change point detection method. Section \ref{sec:multiple_test} provides the change point detection algorithms in different scenarios and their asymptotic theories.
Section \ref{sec:simulation} presents the simulation studies for various signal settings.
Two real data studies are described in Section \ref{sec:data}.
Section \ref{sec:discussion} concludes with a brief discussion.

\section{The mSTEM detection framework}
\label{sec:mSTEM}

\subsection{The kernel smoothed signal}
\label{sec:smoothed_signal}

Consider the model \eqref{eq:signal+noise}. The jump size $a_j$ at $v_j$ is defined as 
\begin{align}\label{eq:a_j}
	a_j = c_{j+1} + k_{j+1}v_j - (c_j + k_jv_j) = (c_{j+1}-c_j) + (k_{j+1}-k_j)v_j, \quad j\geq 1.
\end{align}
By Definition \ref{def:break}, a change point $v_j$ is a Type \RN{1} change point if $a_j=0$ and a Type \RN{2} change point if $a_j\neq 0$. Assume $d=\inf_{j}(v_j-v_{j-1})>0$ that there is a minimal distance $d$ between the neighboring change points. In addition, we assume $k = \inf_j |k_{j+1}-k_j|>0$ at Type \RN{1} change points and $a= \inf_{j}|a_j|>0$ at Type \RN{2} change points, respectively, so that slope changes and jump sizes do not become arbitrarily small.

Denote by $\phi(x)$ and $\Phi(x)$ the pdf and cdf of the standard normal distribution, respectively. 
Let $w_\gamma(t)$ be the Gaussian kernel with compact support $[-c\gamma,c\gamma]$  (we let $c=4$ throughout this paper) and bandwidth $\gamma$, i.e.,
$
w_{\gamma}(t) = \frac{1}{\gamma}\phi\left(\frac{t}{\gamma}\right)\mathbbm{1}\{-c\gamma \leq t \leq c\gamma\}.
$
Convolving the signal-plus-noise process \eqref{eq:signal+noise} with the kernel $w_\gamma(t)$ leads to the generation of a smoothed random process:
\begin{equation}\label{eq:y_gamma}
y_\gamma(t) = w_\gamma(t) * y(t) =
\int_{\mathbb{R}} w_\gamma(t-s) y(s)\,ds = \mu_\gamma(t) + z_\gamma(t),
\end{equation}
where the smoothed signal and smoothed noise are defined, respectively, as
$\mu_\gamma(t) = w_\gamma(t) * \mu(t)$ and $z_\gamma(t) = w_\gamma(t) * z(t)$.
The smoothed noise $z_\gamma(t)$ is assumed to be zero-mean and four-times differentiable.
To avoid the overlap of smoothing between two neighboring change points, we assume that $d = \inf_{j}(v_j - v_{j-1})$ is large enough, specifically, greater than $2c\gamma$.

\subsection{Local extrema for derivatives of the smoothed signal}

For a smooth function $f(t)$, denote by $f^{(\ell)}(t)$ its derivative $\ell$, $\ell \ge 1$, and write by default $f'(t)=f^{(1)}(t)$ and $f''(t)=f^{(2)}(t)$, respectively. We have the following derivatives of the smoothed observed process \eqref{eq:y_gamma},
\begin{equation}
	\label{eq:d1y}
	y^{(\ell)}_\gamma(t) = w^{(\ell)}_\gamma(t) * y(t) =
	\int_{\mathbb{R}} w^{(\ell)}_\gamma(t-s) y(s)\,ds = \mu^{(\ell)}_\gamma(t) + z^{(\ell)}_\gamma(t),
\end{equation}
where the derivatives of the smoothed signal and smoothed noise are 
$\mu^{(\ell)}_\gamma(t) = w^{(\ell)}_\gamma(t) * \mu(t)$ and $z^{(\ell)}_\gamma(t) = w^{(\ell)}_\gamma(t) * z(t)$ for $\ell \ge 1$, respectively.
The following lemmas on the first and second derivatives of the smoothed signal $\mu_{\gamma}(t)$ are useful for characterizing the local extrema (see Figure \ref{fig:illu1}). 
\begin{lemma}
\label{lemma:u1}
The first and second derivatives of $\mu_{\gamma}(t)$ over $(v_{j-1}+c\gamma,v_{j+1}-c\gamma)$ are given, respectively, by
\begin{equation*}\label{eq:d1mu}
\begin{split}
\mu'_{\gamma}(t) &= 
\begin{cases}
 k_j[2\Phi(c) -1],   &\  t\in (v_{j-1}+c\gamma,v_j-c\gamma),\\
 \frac{a_j}{\gamma}\phi(\frac{v_j-t}{\gamma}) + (k_j-k_{j+1})\Phi(\frac{v_j-t}{\gamma}) + (k_j + k_{j+1})\Phi(c) -k_j, &\  t\in (v_j-c\gamma, v_j+c\gamma),\\
 k_{j+1}[2\Phi(c) -1],  &\  t\in (v_j+c\gamma, v_{j+1}-c\gamma);
\end{cases}\\
\mu''_{\gamma}(t) &=
\begin{cases}
\frac{a_j(v_j-t) + (k_{j+1}-k_j)\gamma^2}{\gamma^3}\phi(\frac{v_j-t}{\gamma}), & \quad  t\in (v_j-c\gamma, v_j+c\gamma),\\
0, & \quad \textnormal{otherwise},
\end{cases}
\end{split}
\end{equation*}
where $a_j$ is the jump size defined by \eqref{eq:a_j}.
\end{lemma}


\begin{lemma}\label{lemma:extremum-location}
The local extremum of $\mu'_{\gamma}(t)$ over $(v_j-c\gamma, v_j+c\gamma)$ is achieved at
\begin{subnumcases}{t=}
v_j+\gamma^2 q_j, & \quad \text{\it if } $a_j\neq 0$, \label{eq:d1-1} \\
\text{\it does not exist}, & \quad \text{\it if } $a_j = 0$;  \label{eq:d1-2}
\end{subnumcases}
while the local extremum of $\mu''_{\gamma}(t)$ over $(v_j-c\gamma, v_j+c\gamma)$ is achieved at 
\begin{subnumcases}{t=} v_j + \frac{1}{2}\left(\gamma^2 q_j \pm \gamma\sqrt{\gamma^2q_j^2+4}\right), & \quad \text{\it if } $a_j \neq 0$, \label{eq:d2-1}\\
v_j, & \quad \text{\it if } $a_j = 0$, \label{eq:d2-2}
\end{subnumcases}
where $q_j = \frac{k_{j+1}-k_j}{a_j}$ for $a_j\neq 0$.
\end{lemma}

Recall that $a_j$ is the jump size at the change point $v_j$; therefore, $v_j$ is a Type \RN{1} change point if $a_j=0$, and a Type \RN{2} change point if $a_j\neq 0$. 
Throughout this paper, derivatives are taken with respect to the smoothed signal $\mu_\gamma(t)$, which ensures that derivatives are well defined even in the presence of Type \RN{2} change points, where the underlying signal is not continuous.
Note that $q_j$ is the ratio between the slope change and the jump size. The Lemma \ref{lemma:extremum-location} characterizes the relationship between the peak location of the differentiated smoothed signal and the original location $v_j$, producing the following result.
\begin{proposition}
\label{prop:peak_location}
A Type \RN{1} change point $v_j$ becomes a local extremum in $\mu''_{\gamma}(t)$ precisely at $v_j$ (see \eqref{eq:d2-2});
while a Type \RN{2} change point $v_j$ results in a local extremum in $\mu'_{\gamma}(t)$ at $v_j + \gamma^2q_j$ (see \eqref{eq:d1-1}), which tends to $v_j$ as $q_j \rightarrow 0$. 
\end{proposition}

More specifically, a Type \RN{1} change point $v_j$ becomes a local maximum in $\mu''_{\gamma}(t)$ if $k_{j+1} - k_j >0$ (a local minimum if $k_{j+1} - k_j <0$); and a Type \RN{2} change point $v_j$ results in a local maximum in $\mu'_{\gamma}(t)$ near $v_j$ if $a_j >0$ (a local minimum if $a_j < 0$), regardless of the sign of $k_{j+1} - k_j$. Basically, Proposition \ref{prop:peak_location} shows that the change points would be transformed into peaks in $\mu'_{\gamma}(t)$ or $\mu''_{\gamma}(t)$, providing the main idea for the detection in the pure cases of Scenarios 1 and 2.

\begin{remark}
{\rm Based on the results in the lemma \ref{lemma:extremum-location}, we discuss the idea of detecting the change points of Types \RN{1} and \RN{2} separately in Scenario 3, where both types of change points. This will be shown in more detail in Algorithm \ref{alg:type12} below.

First, note that a Type \RN{1} change point does not generate any local extremum in $\mu'_{\gamma}(t)$ (see \eqref{eq:d1-2}). This property is crucial for detecting Type \RN{2} change points by identifying peaks in $\mu'_{\gamma}(t)$ as stated by Proposition \ref{prop:peak_location} (see step 1 in Algorithm \ref{alg:type12}). On the other hand, as mentioned by \eqref{eq:d2-1}, a Type \RN{2} change point $v_j$ will generate a pair of local maximum and minimum in $\mu''_{\gamma}(t)$, with locations tending to $v_j \pm \gamma$ when $q_j$ is sufficiently small. This will help us remove those local extrema in $\mu''_{\gamma}(t)$ generated from the detected Type \RN{2} change points in the previous step and thus detect the Type \RN{1} change points by identifying peaks in $\mu''_{\gamma}(t)$ as outlined by Proposition \ref{prop:peak_location} (see step 2 in Algorithm \ref{alg:type12}).}
\end{remark}

\subsection{Main ideas}\label{sec:idea}

Figure \ref{fig:illu1} demonstrates the key concept of the proposed change point detection method. The main idea is to transform the problem of change point detection into testing local extrema in the derivatives of the smoothed signal. A candidate peak can arise from either a true change point in the signal region or pure random noise. Multiple testing based on the peak height distribution of the first and second derivatives is then used to identify the significant peaks as the true change points. Although this main idea (or Figure \ref{fig:illu1}) is illustrated in signal $\mu(t)$, it is also applicable to signal-plus-noise $y(t)$ under certain asymptotic conditions such that the signal strength (the size of jump or slope change) and the length of the data sequence are large; see conditions (C1) and (C2) below.

\begin{figure}[!ht]
	\centering
	\includegraphics[width=0.8\textwidth]{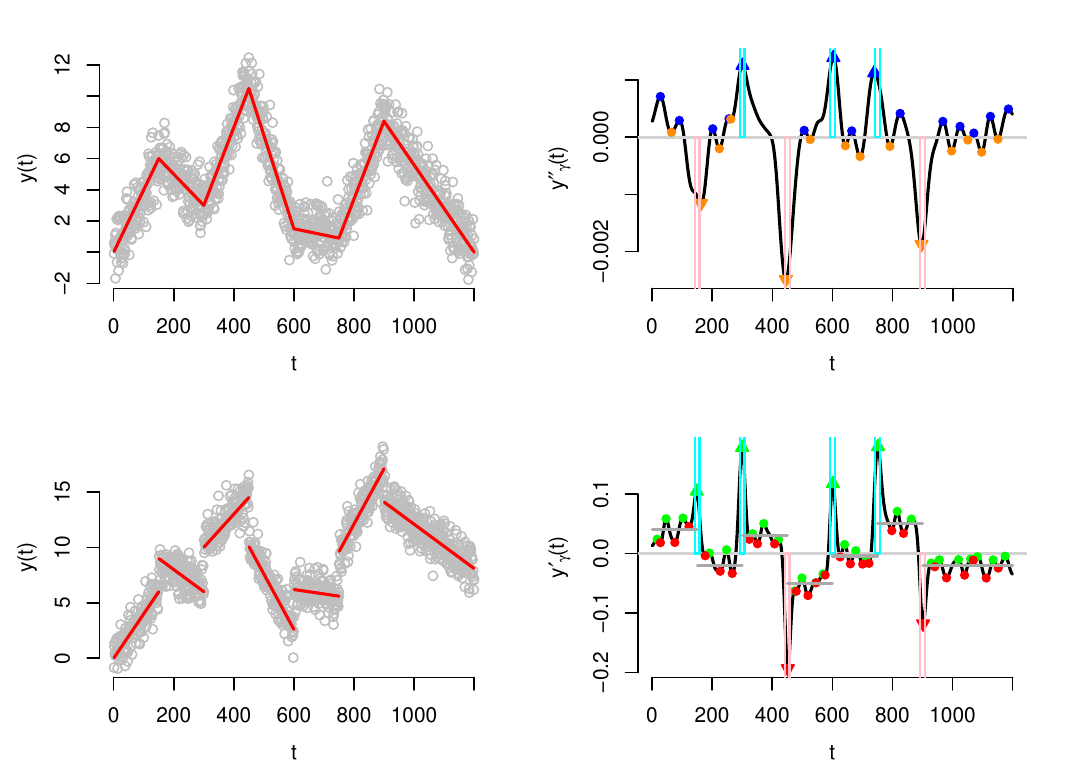} 
	\caption{\footnotesize Process of Type \RN{1} and Type \RN{2} change point detection.
	The two plots on the left panel show the data (grey circles) with six Type \RN{1} change points and the data with six Type \RN{2} change points respectively. The red lines indicate piecewise linear signals $\mu(t)$. 
	The two plots on the right panels show the second and first derivatives of the smoothed data ($\gamma=10$), respectively.	
	In $y'_{\gamma}(t)$, local maxima and local minima are represented by green and red dots respectively. In $y''_{\gamma}(t)$, they are represented by blue and orange dots respectively. Triangles indicate the significant local extrema at significant level $\alpha=0.05$. 
	The cyan and pink bars show the location tolerance intervals $(v_j-b, v_j+b)$ with $b=5$ for the true change points. 
	The grey horizontal lines (lower right plot) indicate piecewise slopes of the signal, serving as baselines in the testing of $y'_{\gamma}(t)$. In these two testings, 
    all change points of both Type \RN{1} and Type \RN{2} are detected correctly without any false discovery.}
	\label{fig:illu2_1}
\end{figure}

\begin{figure}[!ht]
	\centering
	\includegraphics[width = 0.9\textwidth]{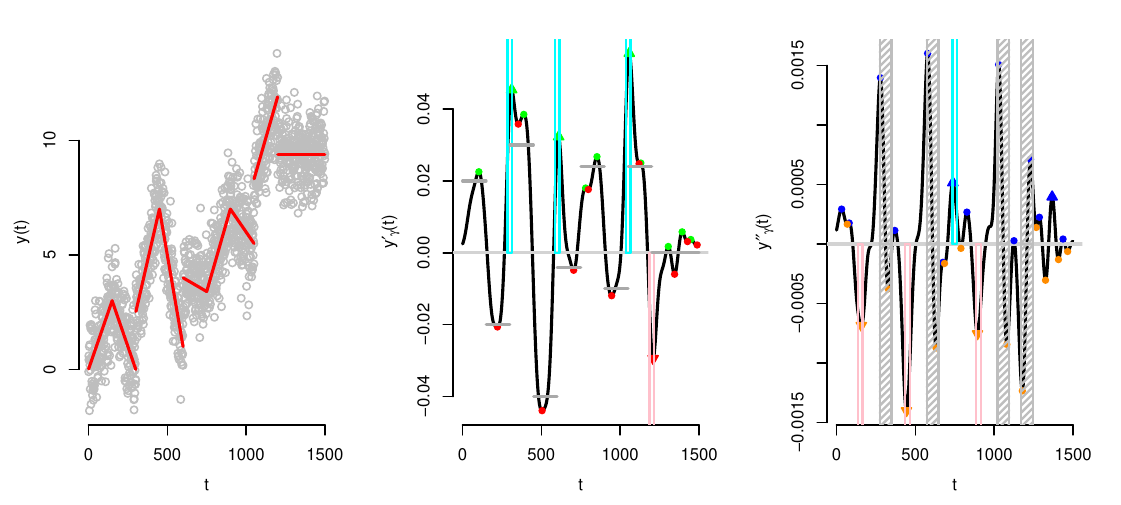} 
	\caption{\footnotesize Process of mixture of Type \RN{1} and Type \RN{2} change point detection.
	The same graphical symbols and colors as in Figure \ref{fig:illu2_1} are utilized here.
	The left plot (grey circles) shows the data with four Type \RN{1} and four Type \RN{2} change points. 
	The middle and right plots show the detection of Type \RN{2} change points in $y'_{\gamma}(t)$ and the detection of Type \RN{1} change points in $y''_{\gamma}(t)$, respectively.
Type \RN{2} change points are first detected in the first derivative $y'_{\gamma}(t)$, then the local extrema in the second derivative $y''_{\gamma}(t)$ generated by the detected Type \RN{2} change points are removed (represented by grey shaded bars and each bar is $[v_j-2\gamma,v_j+2\gamma]$).
 After this filtering process, the remaining significant local extrema in $y''_{\gamma}(t)$ are identified as Type \RN{1} change points. In this testing, all Type \RN{1} and Type \RN{2} change points are detected correctly without any false discovery.}
	\label{fig:illu2_2}
\end{figure}
 
To further illustrate the main idea, especially the proposed algorithm, toy examples are presented in Figures \ref{fig:illu2_1} and \ref{fig:illu2_2}. 
 In Scenario 1 (pure Type \RN{1} change points), a Type \RN{1} change point generates a peak exactly at its location in the second derivative $\mu''_\gamma(t)$, as indicated in \eqref{eq:d2-2}. 
 In Scenario 2 (pure Type \RN{2} change points), a Type \RN{2} change point produces a peak in the first derivative $\mu'_\gamma(t)$ around its location, as described in \eqref{eq:d1-1}. By subtracting the corresponding baseline (piecewise slope at $v_j$) from the peaks in $y'_{\gamma}(t)$, the problem reduces to a standard peak detection problem. In Scenario 3, a combination of Type \RN{1} and Type \RN{2} change points is considered. The Type \RN{2} change points are first identified based on the peaks in $y'_{\gamma}(t)$, and then the Type \RN{1} change points are detected as significant peaks in $y''_{\gamma}(t)$ after cutting off the Type \RN{2} change points. 

Our proposed method, called {mSTEM} (\textbf{m}odified \textbf{S}moothing and \textbf{TE}sting of \textbf{M}axima/\textbf{M}inima), consists of the following steps.
1. Differential kernel smoothing: to transform change points into local maxima or minima (illustrated in Figure \ref{fig:illu1}), and can meanwhile increase SNR;
2. Candidate peaks: to find local extrema of either the first or second derivative of the smoothed observed data;
3. P-values: to compute the p-value of each local maximum or minimum under the null hypothesis of no change point (no signal) in a local neighborhood;
4. Multiple testing: to apply a multiple testing procedure (Benjamini-Hochberg procedure) to the set of local maxima and minima; a change point is claimed to be detected if the p-value is significant.

\section{Multiple change point detection for linear models}
\label{sec:multiple_test}

Suppose that we observe $y(t)$ defined by \eqref{eq:signal+noise} with $J$ change points in the line of length $L$ centered at the origin, which is denoted by $U(L)=(-L/2,L/2)$.
We define \emph{signal region} as $\mathbb{S}_1 =  \cup_{j=1}^{J}S_j = \cup_{j=1}^J (v_j-b, v_j+b)$, which is the region of interest where true change points are expected to occur. It is defined as the union of intervals centered around each true change point $v_j$. The \emph{null region} is the complement of the signal region within $U(L)$ given by $\mathbb{S}_0 = U(L) \setminus \mathbb{S}_1$, where true change points are not expected to occur.

\subsection{Pure Type \RN{1} change point detection}
\label{sec:type1}

We first consider the case of pure type \RN{1} change points, which do not generate any peaks in $\mu_\gamma'(t)$ by \eqref{eq:d1-2}. Applying Proposition \ref{prop:peak_location}, it is appropriate to detect the Type \RN{1} change points by detecting peaks in $y_\gamma''(t)$.
Following the mSTEM procedure proposed for the detection of change points in Section \ref{sec:idea}, we present the mSTEM algorithm for detecting Type \RN{1} change points.

\begin{algo}[mSTEM algorithm for Type \RN{1} change point detection]
\label{alg:type1}
\hfill\par\noindent
1. {\em Differential kernel smoothing}:
		Obtain the process $y''_\gamma(t)$ in \eqref{eq:d1y} by convolution of $y(t)$ with the kernel derivative $w''_\gamma(t)$.
        
\noindent 2. {\em Candidate peaks}:
		Find the set of local maxima and minima of $y''_\gamma(t)$ in $U(L)$, denoted by $\tilde{T}_{\RN{1}} = \tilde{T}^{+}_{\RN{1}} \cup \tilde{T}^{-}_{\RN{1}}$, where 
  $
  \tilde{T}^{+}_{\RN{1}} = \{ t \in U(L): y^{(3)}_{\gamma}(t) = 0, \  y^{(4)}_{\gamma}(t) < 0\},\  \tilde{T}^{-}_{\RN{1}} = \{ t \in U(L): y^{(3)}_{\gamma}(t) = 0, \  y^{(4)}_{\gamma}(t) > 0\}.
  $
  Note that $\tilde{T}^{+}_{\RN{1}}$ and $\tilde{T}^{-}_{\RN{1}}$ are the sets of local maxima and minima in $y''_{\gamma}(t)$.
  
\noindent 3. {\em P-values}:
		For each $t \in \tilde{T}^{+}_{\RN{1}}$, compute the p-value $p_{\RN{1}}(t)$ for testing the hypotheses
		$
			\mathcal{H}_{0}(t): \{\mu''(s) = 0\  \text{ for all } \ s \in (t-b,t+b) \} \; {\rm vs.} 
        $ 
        $
			\mathcal{H}_{A}(t): \{\mu'(s+) > \mu'(s-) \ \text{ for some } \ s \in (t-b,t+b) \};
        $ 
		and for each $t \in \tilde{T}^{-}_{\RN{1}}$, compute the p-value $p_{\RN{1}}(t)$ for the hypotheses
		$
			\mathcal{H}_{0}(t): \ \{\mu''(s) = 0 \ \text{ for all } \ s \in (t-b,t+b) \} \ {\rm vs.} 
        $
        $
			\mathcal{H}_{A}(t): \ \{\mu'(s+) < \mu'(s-) \ \text{ for some } \ s \in (t-b,t+b) \},
        $
		where $\mu'(s+)=\lim_{x\to s+} \mu'(x)$, $\mu'(s-)=\lim_{x\to s-} \mu'(x)$ and $b>0$ is an appropriate location tolerance parameter.	
        
\noindent 4. {\em Multiple testing}: Apply the Benjamini-Hochberg (BH) multiple testing procedure on the set of p-values $\{p_{\RN{1}}(t), \, t \in \tilde{T}_{\RN{1}}\}$, defined in Section \ref{sec:p-value} below, and declare all significant local extrema whose p values are lower than the significance threshold.
\end{algo}

%
\subsubsection{P-value calculation}\label{sec:p-value}

The p-value in step 3 of Algorithm \ref{alg:type1} is given by
\begin{equation}
\label{eq:pval}
p_{\RN{1}}(t) =\begin{cases}
F_{z''_{\gamma}}(y''_\gamma(t)) \quad &t \in \tilde{T}^{+}_{\RN{1}}, \\
F_{z''_{\gamma}}(-y''_\gamma(t)) \quad &t \in \tilde{T}^{-}_{\RN{1}},
\end{cases}
\end{equation}
where $F_{z''_{\gamma}}(u)$ (defined in Section \ref{sec:gau-auto} below) is the conditional right tail probability of $z''_\gamma(t)$ at the local maximum $t \in \tilde{T}_{\RN{1}}^+$, evaluated under the null model $\mu_{\gamma}''(s) = 0, \forall s\in (t-b, t+b)$, i.e.,
\begin{equation}
\label{eq:F2x}
F_{z''_{\gamma}}(u) = P\left(z''_\gamma(t) > u \big| \text{ $t$ is a local maximum of $z''_\gamma(t)$}\right).
\end{equation}
The second line in \eqref{eq:pval} is due to the symmetry property that 
$P\left(z''_\gamma(t) < u  \big| \text{ $t$ is a local minimum of $z''_\gamma$}\right)$ 
becomes  $P\left(-z''_\gamma(t) > -u \big| \text{ $t$ is a local maximum of $-z''_\gamma$}\right) =  F_{z''_{\gamma}}(-u)$ since $z''_\gamma(t)$ and $-z''_\gamma(t)$ have the same distribution.

\subsubsection{Error and power definitions}

We define $\tilde{T}_{\RN{1}}(u) = \tilde{T}^{+}_{\RN{1}}(u) \cup \tilde{T}^{-}_{\RN{1}}(u)$, where
$
\tilde{T}^{+}_{\RN{1}}(u) = \left\{ t \in U(L): y_\gamma''(t)>u, \ y^{(3)}_{\gamma}(t) = 0, \  y^{(4)}_{\gamma}(t) < 0\right\}
$
and 
$
\tilde{T}^{-}_{I}(u) = \left\{ t \in U(L): y_\gamma''(t)<-u, \ y^{(3)}_{\gamma}(t) = 0, \  y^{(4)}_{\gamma}(t) > 0\right\}.
$

It is seen that $\tilde{T}^{+}_{\RN{1}}(u)$ and $\tilde{T}^-_{1}(u)$ are the sets of local maxima of $y_\gamma''(t)$  above $u$ and the sets of local minima of $y_\gamma''(t)$ below $-u$, respectively. The number of totally and falsely detected change points at threshold $u$ is defined respectively as
$R_{\RN{1}}(u) = \#\{t\in \tilde{T}_{\RN{1}}(u)\}$ and $V_{\RN{1}}(u) =\#\{t\in \tilde{T}_{\RN{1}}(u)\cap \mathbb{S}_0\}$.
Both are defined as zero if $\tilde{T}_{\RN{1}}(u)$ is empty. The FDR at threshold $u$ is defined as the expected proportion of falsely detected change points
$
{\rm FDR}_{\RN{1}}(u) = {\rm E}\left\{ \frac{V_{\RN{1}}(u)}{R_{\RN{1}}(u)\vee 1} \right\}.
$

Denote by $\mathcal{I}_{\RN{1}}^+$ and $\mathcal{I}_{\RN{1}}^-$ the collections of indices $j$ corresponding to the increasing and decreasing (in slope change) Type \RN{1} change points, respectively. The power is defined as 
\begin{equation}
\label{eq:power}
\begin{split}
{\rm Power}_{\RN{1}}(u) &= \frac{1}{J}\sum_{j=1}^J P\left(\tilde{T}_{\RN{1}}(u)\cap S_j \ne \emptyset \right) \\
&= \frac{1}{J}\Bigg[\sum_{j\in \mathcal{I}_{\RN{1}}^+} P\left(\tilde{T}_{\RN{1}}^+(u)\cap S_j \ne \emptyset \right)  + \sum_{j\in \mathcal{I}_{\RN{1}}^-} P\left(\tilde{T}_{\RN{1}}^-(u)\cap S_j \ne \emptyset \right)\Bigg],
\end{split}
\end{equation}
where $P(\cdot)$ is the probability of detecting the change point $v_j$ within the signal region $S_j=(v_j-b, v_j+b)$. 
Throughout this paper, when considering the local extrema of the second derivative of the process over the signal region $S_j$ or the smoothed signal region $S_{j,\gamma}=(v_j - c\gamma, v_j + c\gamma)$ around $j$, the local extrema are regarded as local maxima if $j\in \mathcal{I}_{\RN{1}}^+$ and as local minima if $j\in \mathcal{I}_{\RN{1}}^-$.

\subsubsection{Asymptotic FDR control and power consistency}

Denote by $E[\tilde{m}_{z''_{\gamma}}(U(1))]$ and $E[\tilde{m}_{z''_{\gamma}}(U(1),u)]$ the expected number of local extrema, and the expected number of both local maxima above the level $u$ and local minima below $-u$ of $z''_{\gamma}(t)$ in the unit interval $U(1) = (-1/2, 1/2)$, respectively. Note that, by symmetry, the expected number of local minima of $z''_{\gamma}(t)$ below the level $-u$ is equal to the expected number of local maxima of $z''_{\gamma}(t)$ above $u$, which is $E[\tilde{m}_{z''_{\gamma}}(U(1),u)]/2$.
Similarly, we define $E[\tilde{m}_{z'_{\gamma}}(U(1))]$ and $E[\tilde{m}_{z'_{\gamma}}(U(1),u)]$ for $z'_{\gamma}(t)$ on $U(1)$.

Recall that the BH procedure applied in step 4 of Algorithm \ref{alg:type1} is defined as follows. Let $m_{\RN{1}}$ be the size of the set $\tilde{T}_{\RN{1}}$. For a fixed $\alpha \in (0,1)$, let $\ell$ be the largest index for which the $i$th smallest p-value is less than $i\alpha/m_{\RN{1}}$. Then the null hypothesis $\mathcal{H}_0(t)$ at $t \in \tilde{T}_{\RN{1}}$ is rejected if
\begin{equation}
\label{eq:thresh-BH-random1}
p_{\RN{1}}(t) < \frac{\ell\alpha}{m_{\RN{1}}} \iff
\begin{cases}
y''_\gamma(t) > \tilde{u}_{\RN{1}} = F_{z''_{\gamma}}^{-1} \left(\frac{\ell\alpha}{m_{\RN{1}}}\right) & \text{ if } t\in \tilde{T}^+_{\RN{1}}, \\ y''_\gamma(t) < -\tilde{u}_{\RN{1}} = -F_{z''_{\gamma}}^{-1} \left(\frac{\ell\alpha}{m_{\RN{1}}}\right) & \text{ if } t\in \tilde{T}^-_{\RN{1}}, \end{cases}
\end{equation}
where $\ell\alpha/{m}_{\RN{1}}$ is defined as 1 if $m_{\RN{1}}=0$. Note that $\tilde{T}_{\RN{1}}$ and therefore $m_{\RN{1}}$ and $\tilde{u}_{\RN{1}}$ are also random, which is different from the usual BH procedure. We define the FDR in this BH procedure as
${\rm FDR}_{{\RN{1}}, {\rm BH}} = E\left\{ \frac{V_{\RN{1}}(\tilde{u}_{\RN{1}})}{R_{\RN{1}}(\tilde{u}_{\RN{1}})\vee1} \right\}$.
The expectation is taken over all possible realizations of the random threshold $\tilde{u}_{\RN{1}}$.
In contrast to the standard BH procedure,  the number of p-values, $m_{\RN{1}}$, is random.
Similarly, the corresponding power, ${\rm Power}_{\RN{1},\rm BH}$, is defined as \eqref{eq:power} with $u$ replaced by $\tilde{u}_{\RN{1}}$.

For the asymptotic theories of Type \RN{1} change point detection, we make the assumption:

\noindent (C1). $L \rightarrow \infty$, $k = \inf_j |k_{j+1} - k_j| \rightarrow \infty$ and $(\log L)/k^2 \rightarrow 0$.

\begin{theorem}\label{thm:type1_fdr}
	Suppose that $y(t)$ contains only Type \RN{1} change points, condition (C1) holds and $J/L \to A >0 $ as $L\rightarrow \infty$. 
    
\noindent (i) If Algorithm \ref{alg:type1} is applied with a fixed threshold $u$, then
	\begin{align}
	{\rm FDR}_{\RN{1}}(u) \rightarrow  \frac{E[\tilde{m}_{z''_{\gamma}}(U(1),u)](1-2c\gamma A)}{E[\tilde{m}_{z''_{\gamma}}(U(1),u)](1-2c\gamma A) + A}\quad \text{and} \quad 
    {\rm Power}_{\RN{1}}(u) \rightarrow 1.  
	\end{align}
    
\noindent (ii) If Algorithm \ref{alg:type1} is applied with the random threshold $\tilde{u}_{\RN{1}}$ \eqref{eq:thresh-BH-random1}, then
	\begin{align}
{\rm FDR}_{\RN{1},\rm BH}
	\rightarrow \alpha \frac{E[\tilde{m}_{z''_{\gamma}}(U(1))](1-2c\gamma A)}{E[\tilde{m}_{z''_{\gamma}}(U(1))](1-2c\gamma A) + A} \quad \text{and}\quad
    {\rm Power}_{\RN{1},\rm BH} \rightarrow 1.  
	\end{align} 
\end{theorem}

Under the asymptotic condition (C1), the BH procedure with the random threshold $\tilde{u}_{\RN{1}}$ in \eqref{eq:thresh-BH-random1} has asymptotically the same error control properties as if the threshold were deterministic and given by
$u_{\RN{1}}^* = F_{z''_{\gamma}}^{-1}(\frac{\alpha A}{A + E[\tilde{m}_{z''_{\gamma}}(U(1))](1-2c\gamma A) (1-\alpha)})$ (see Section S6 in the Supplementary Materials).

By the definition of $d$, we have $Jd<L$ or $d<L/J \to 1/A$. Meanwhile, it is assumed $d = \inf_j (v_j - v_{j-1}) > 2c\gamma$, thus we have $1/A>2c\gamma$ and hence $1 - 2 c\gamma A > 0$.
Note that $2c\gamma J$ represents the length of smoothed signal regions and $(L - 2c\gamma J)$ is the length of smoothed null regions, which are the complements of smoothed signal regions.
Part $(i)$ of Theorem \ref{thm:type1_fdr} shows that FDR converges to the ratio of the number of local extrema generated by random noise to the total number of local extrema, which includes those induced by random noise and the true change points. Part $(ii)$ guaranties that FDR is asymptotically controlled at a prespecified general significant level $\alpha$.

\begin{remark}
Condition (C1) assumes that the signal strength $k$ increases with the search space $L$ to remove excess error caused by the smoothed signal spreading into neighboring null regions, which are the complements of signal regions, thereby enabling asymptotic error control.
This assumption is not restrictive, as the search space grows exponentially faster than the signal strength. 
Viewing the data as pointwise test statistics, this assumption is analogous to the conditions required for the consistency of model selection in high-dimensional regression. Specifically, $(\log p)/n \rightarrow 0$ where $p$ is the number of features and $n$ is the sample size.
Additionally, condition (C1) is easy to state but is much stronger than necessary. It suffices for $(\log L)/k^2$ to be bounded by a constant that depends on the third or fourth derivatives of the smoothed noise (see Remark S1 in the Supplementary Materials). The numerical studies in Section \ref{sec:simulation} further demonstrate that FDR control and power consistency are maintained even in non-asymptotic settings with moderate slope changes.
\end{remark}
\subsection{Pure Type \RN{2} change point detection}
\label{sec:type2}

In the detection of Type \RN{1} breaks, taking the second derivative, the piecewise linear parts of $\mu(t)$ become 0 in $\mu''_{\gamma}(t)$ over $\mathbb{R}\setminus \mathbb{S}_{1,\gamma}$, where $\mathbb{S}_{1,\gamma} = \cup_{j=1}^J (v_j - c\gamma, v_j + c\gamma)$ is the \emph{smoothed signal region}  (see the intervals indicated by the blue points in Figure \ref{fig:illu1}). However, in the detection of Type \RN{2} breaks, the piecewise linear parts become piecewise constants (the slopes) in $\mu'_{\gamma}(t)$ over $\mathbb{R}\setminus \mathbb{S}_{1,\gamma}$. To detect Type \RN{2} breaks, the null hypothesis (no signal/jump) $\mu'_{\gamma}(t) = k(t)$, where $k(t)$ is the corresponding piecewise slope at $t$, will be tested. Hence, it is crucial to estimate the piecewise slopes around the change points $v_j$. 

\subsubsection{Piecewise slopes estimate}
\label{sec:slope}

The main idea of estimating piecewise slopes is to divide the data sequence into segments and estimate the slopes within each segment using linear regression. 
These segments are determined based on the presence of pairwise local maxima and minima, which are located near $v_j\pm \gamma$ around the change points in the second derivative $\mu''_{\gamma}(t)$ (see Figure \ref{fig:illu1} (f)).

To ensure that all true Type \RN{2} breaks are captured and to provide a better estimate of the piecewise slopes, Algorithm \ref{alg:type1} is applied with a larger significance level, such as 0.1. This helps in obtaining non-conservative initial estimators for Type \RN{2} change points based on the peaks detected in $y''_{\gamma}(t)$. 
Then, within each segment, the piecewise slope is estimated using a robust regression model \citep{huber2004robust} to mitigate the influence of the change points.

Once the piecewise slopes are obtained, we will apply the following algorithm, which is based on the local extrema of the first derivative of the smoothed data, to detect Type \RN{2} change points.

 \begin{algo}[mSTEM algorithm for Type \RN{2} change point detection]
 	\label{alg:type2}
 	\hfill\par\noindent
1. {\em Differential kernel smoothing}:
 		Obtain the process $y'_{\gamma}(t)$ in \eqref{eq:d1y} by convolution of $y(t)$ with the kernel derivative $w'_\gamma(t)$.
        
\noindent 2. {\em Candidate peaks}:
 		Find the set of local maxima and minima of $y'_\gamma(t)$ in $U(L)$, denoted by $\tilde{T}_{\RN{2}} = \tilde{T}^{+}_{\RN{2}} \cup \tilde{T}^{-}_{\RN{2}}$, where
 		\begin{equation*}
 				\tilde{T}^{+}_{\RN{2}} = \left\{ t \in U(L): y''_{\gamma}(t) = 0, \  y^{(3)}_{\gamma}(t) < 0\right\},\ 
 				\tilde{T}^{-}_{\RN{2}} = \left\{ t \in U(L): y''_{\gamma}(t) = 0, \  y^{(3)}_{\gamma}(t) > 0\right\}.
 		\end{equation*}
\noindent 3. {\em P-values}:
 		For each $t \in \tilde{T}^{+}_{\RN{2}}$, compute the p-value $p_{\RN{2}}(t)$ for testing the hypotheses
 		$$
 		\begin{aligned}
 			\mathcal{H}_{0}(t):& \ \{\mu'(s) =k(s) \quad \text{for all} \quad s \in (t-b,t+b) \} \quad {\rm vs.} \\
 			\mathcal{H}_{A}(t):& \ \{\mu(s+)>\mu(s-)  \quad \text{for some} \quad s \in (t-b,t+b) \},
 		\end{aligned}
 		$$
 		where $k(s)$ is the piecewise slope at $s$, $\mu(s+)=\lim_{x\to s+} \mu(x)$, $\mu(s-)=\lim_{x\to s-} \mu(x)$ and $b>0$ is an appropriate location tolerance parameter.
 	    For each $t \in \tilde{T}^{-}_{\RN{2}}$, compute the p-value $p_{\RN{2}}(t)$ for testing the hypotheses
 	$
         \mathcal{H}_{0}(t): \ \{\mu'(s) = k(s) \ \text{ for all } \ s \in (t-b,t+b) \} \ {\rm vs.}
        $		
        $
        \mathcal{H}_{A}(t): \ \{\mu(s+) < \mu(s-) \ \text{ for some } \ s \in (t-b,t+b) \}.
        $
        
\noindent 4. {\em Multiple testing}: Apply the BH procedure on the set of p-values $\{p_{\RN{2}}(t), \, t \in \tilde{T}_{\RN{2}}\}$, and declare significant all local extrema whose p-values are smaller than the significance threshold.
 \end{algo}

Similarly to \eqref{eq:pval} and \eqref{eq:F2x}, the p-value in step 3 of Algorithm \ref{alg:type2} is given by
\begin{equation*}
p_{\RN{2}}(t) =\begin{cases}
F_{z'_{\gamma}}(y'_\gamma(t)-k(t)) \quad &t \in \tilde{T}^{+}_{\RN{2}}, \\
F_{z'_{\gamma}}(-(y'_\gamma(t)-k(t))) \quad &t \in \tilde{T}^{-}_{\RN{2}}.
\end{cases}
\end{equation*}
Here $F_{z'_{\gamma}}(u)$ is the right tail probability of $z'_\gamma(t)$ at the local maximum $t \in \tilde{T}_{\RN{2}}^+$, as defined in Section \ref{sec:gau-auto} below, which is $F_{z'_{\gamma}}(u) = P\left(z'_\gamma(t) > u \big| \text{ $t$ is a local maximum of $z'_\gamma(t)$}\right)$.

\subsubsection{Error and power definitions}

We define $\tilde{T}_{\RN{2}}(u) = \tilde{T}^{+}_{\RN{2}}(u) \cup \tilde{T}^{-}_{\RN{2}}(u)$, where
\begin{align*}
& \tilde{T}^{+}_{\RN{2}}(u) = \left\{ t \in U(L): y_\gamma'(t) - k(t) > u, \ y''_{\gamma}(t) = 0, \  y^{(3)}_{\gamma}(t) < 0\right\},\\
& \tilde{T}^{-}_{\RN{2}}(u) = \left\{ t \in U(L): y_\gamma'(t) - k(t) < -u, \ y''_{\gamma}(t) = 0, \  y^{(3)}_{\gamma}(t) > 0\right\}.
\end{align*}

The number of totally and falsely detected change points and FDR at threshold $u$ are defined as  
$
R_{\RN{2}}(u) = \#\{t\in \tilde{T}_{\RN{2}}(u)\}, \; V_{\RN{2}}(u) =\#\{t\in \tilde{T}_{\RN{2}}(u)\cap \mathbb{S}_0\}, \;  {\rm FDR}_{\RN{2}}(u) = {\rm E}\left\{ \frac{V_{\RN{2}}(u)}{R_{\RN{2}}(u)\vee 1} \right\} ,
$ respectively.

Denote by $\mathcal{I}_{\RN{2}}^+$ and $\mathcal{I}_{\RN{2}}^-$ the sets of indices $j$ corresponding to increasing and decreasing (in jump) Type \RN{2} change points $v_j$, respectively. The power is defined as 
\begin{equation}
\label{eq:power2}
\begin{split}
{\rm Power}_{\RN{2}}(u) &= \frac{1}{J}\sum_{j=1}^J P\left(\tilde{T}_{\RN{2}}(u)\cap S_j \ne \emptyset \right) \\
&= \frac{1}{J}\Bigg[\sum_{j\in \mathcal{I}_{\RN{2}}^+} P\left(\tilde{T}_{\RN{2}}^+(u)\cap S_j \ne \emptyset \right)  + \sum_{j\in \mathcal{I}_{\RN{2}}^-} P\left(\tilde{T}_{\RN{2}}^-(u)\cap S_j \ne \emptyset \right)\Bigg].
\end{split}
\end{equation}
Throughout this paper, when considering local extrema over the signal region $S_j=(v_j-b, v_j+b)$ or the smoothed signal region $S_{j,\gamma}=(v_j - c\gamma, v_j + c\gamma)$ around $j$, the local extrema of the derivative of the process are regarded as local maxima if $j\in \mathcal{I}_{\RN{2}}^+$ and as local minima if $j\in \mathcal{I}_{\RN{2}}^-$, respectively.

\subsubsection{Asymptotic FDR control and power consistency}

Suppose the BH procedure is applied in step 4 of Algorithm \ref{alg:type2} as follows. Define $m_{\RN{2}}$ as the size of set $\tilde{T}_{\RN{2}}$. For a fixed $\alpha \in (0,1)$, let $\ell$ be the largest index for which the $i$th smallest p-value is less than $i\alpha/m_{\RN{2}}$. Then the null hypothesis $\mathcal{H}_0(t)$ at $t \in \tilde{T}_{\RN{2}}$ is rejected if
\begin{equation}
\label{eq:thresh-BH-random2}
p_{\RN{2}}(t) < \frac{\ell\alpha}{m_{\RN{2}}} \iff
\begin{cases}
y'_\gamma(t) - k(t) > \tilde{u}_{\RN{2}} = F_{z'_{\gamma}}^{-1} \left(\frac{\ell\alpha}{m_{\RN{2}}}\right) & \text{ if } t\in \tilde{T}^+_{\RN{2}}, \\ y'_\gamma(t) - k(t) < -\tilde{u}_{\RN{2}} = -F_{z'_{\gamma}}^{-1} \left(\frac{\ell\alpha}{m_{\RN{2}}}\right) & \text{ if } t\in \tilde{T}^-_{\RN{2}}, \end{cases}
\end{equation}
where $F_{z'_{\gamma}}$ is defined in \eqref{eq:F2x} with $z''_{\gamma}$ replaced by $z'_{\gamma}$, and $\ell\alpha/m_{\RN{2}}$ is defined as 1 if $m_{\RN{2}}=0$. Since $\tilde{u}_{\RN{2}}$ is random, we define the FDR in such a BH procedure as
${\rm FDR}_{{\RN{2}}, {\rm BH}} = E\left\{ \frac{V_{\RN{2}}(\tilde{u}_{\RN{2}})}{R_{\RN{2}}(\tilde{u}_{\RN{2}})\vee1} \right\}$. 
Similarly, the corresponding power, denoted by ${\rm Power}_{\RN{2},\rm BH}$, is defined as \eqref{eq:power} with $u$ replaced by $\tilde{u}_{\RN{2}}$.

To establish asymptotic theories for Type \RN{2} change point detection, we make the assumption:

\noindent (C2).  $a = \inf_j |a_j| \rightarrow \infty$, $q = \sup_{j} |\frac{k_{j+1}-k_j}{a_j}| \to 0$ and $\frac{\log L}{a^2} \rightarrow 0$ as $L\to \infty$.

\begin{theorem}\label{thm:type2_fdr}
	Suppose $y(t)$ contains only Type \RN{2} change points, the condition (C2) holds and $J/L \to A >0 $ as $L\rightarrow \infty$. 

\noindent (i) If Algorithm \ref{alg:type2} is applied with a fixed threshold $u$, then
	\begin{align*}
	&{\rm FDR}_{\RN{2}}(u) \rightarrow  \frac{E[\tilde{m}_{z'_{\gamma}}(U(1),u)](1-2c\gamma A)}{E[\tilde{m}_{z'_{\gamma}}(U(1),u)](1-2c\gamma A) + A} \quad \text{and} \quad {\rm Power}_{\RN{2}}(u) \rightarrow 1. 
	\end{align*}
    
\noindent (ii) If Algorithm \ref{alg:type2} is applied with a random threshold $\tilde{u}_{\RN{2}}$ \eqref{eq:thresh-BH-random2}, then
	\begin{align*}
	& {\rm FDR}_{\RN{2},\rm BH}
	 \rightarrow \alpha \frac{E[\tilde{m}_{z'_{\gamma}}(U(1))](1-2c\gamma A)}{E[\tilde{m}_{z'_{\gamma}}(U(1))](1-2c\gamma A) + A}\quad \text{and} \quad {\rm Power}_{\RN{2},\rm BH} \rightarrow 1.
	\end{align*} 
\end{theorem}

Condition (C2) is similar to condition (C1) but focuses on the jump size. It requires the jump size to dominate the slope changes, ensuring a clear distinction between Type I and Type II change points. Similarly, the limit of $(\log L)/a^2$ needs only to be bounded by a constant that depends on the second and third derivatives of the smoothed signal (see Lemma S3 and Remark S1 in Supplementary Materials).

\subsection{Mixture of Type \RN{1} and Type \RN{2} change point detection}

We have shown that pure Type \RN{1} and Type \RN{2} change points can be detected through peak detection in the second derivative $y''_{\gamma}(t)$ (see Algorithm \ref{alg:type1}) and the first derivative $y'_{\gamma}(t)$ (see Algorithm \ref{alg:type2}), respectively.
However, it is very common for a real signal to contain both Type \RN{1} and Type \RN{2} change points.  Distinguishing between Type \RN{1} and Type \RN{2} change points is a key challenge when managing signals comprising a combination of these types. Our strategy involves initially identifying Type \RN{2} change points via peak detection within the first derivative $y'_{\gamma}(t)$, and then detecting Type \RN{1} change points by excluding the locations of Type \RN{2} change points from consideration in the second derivative $y''_{\gamma}(t)$. 
The rationale behind this approach is as follows:
Type \RN{2} change points generate peaks in the first derivative $y'_{\gamma}(t)$, while Type \RN{1} change points do not. By detecting significant peaks in $y'_{\gamma}(t)$ using Algorithm \ref{alg:type2}, we can identify the locations of Type \RN{2} change points.
Once the Type \RN{2} change points are identified, we can focus on detecting Type \RN{1} change points. Note that both Type \RN{1} and Type \RN{2} change points can produce peaks in the second derivative $y''_{\gamma}(t)$. However, some of these peaks may be generated by Type \RN{2} change points, and we want to exclude them from our Type \RN{1} detection.


To distinguish between Type \RN{1} and Type \RN{2} change points in the mixture case, we remove the Type \RN{2} change points detected in the first step. This means that any peaks in the second derivative $y''_{\gamma}(t)$ that coincide with the locations of Type \RN{2} change points are excluded from the set of potential Type \RN{1} change points. The remaining peaks in $y''_{\gamma}(t)$, which are not associated with Type \RN{2} change points, are more likely to be generated by Type \RN{1} change points.


\begin{algo}[mSTEM algorithm for mixture of Type \RN{1} and \RN{2} change point detection]
	\label{alg:type12}
	\hfill\par\noindent
    
\noindent 1. {\em Detect Type \RN{2} breaks}:
		Perform Algorithm \ref{alg:type2} to obtain estimates of Type \RN{2} breaks, denoted as $\hat{v}_{i} \in \tilde{T}_\RN{2}$ for $i=1,\dots, R_\RN{2}$.
		A larger $\gamma$ in this step is suggested to achieve a better estimate of Type \RN{2} change points and fewer false discoveries. Because a larger $\gamma$ generally results in a higher SNR (see Section \ref{sec:snr}).
        
\noindent 2. {\em Candidate Type \RN{1} peaks}:
		Find the set of local maxima and minima of $y''_\gamma(t)$ in $U(L)$, denoted by $\tilde{T}_{\RN{1}} = \tilde{T}^{+}_{\RN{1}} \cup \tilde{T}^{-}_{\RN{1}}$.
	    As $\tilde{T}_{\RN{1}}$ contains local extrema generated by both Type \RN{1} and Type \RN{2} breaks, it is necessary to remove the peaks generated by Type \RN{2} breaks that are detected in step 1.
     Let $\mathbb{S}^\ast_\RN{2} = \cup_{i=1}^{R_\RN{2}}(\hat{v}_{i}-2\gamma,\hat{v}_{i}+2\gamma)$ be the removed region, then the set of candidate peaks of Type \RN{1} breaks is defined as $\tilde{T}_{\RN{1} \setminus \RN{2}}=\tilde{T}^{+}_{\RN{1} \setminus \RN{2}}\cup\tilde{T}^{-}_{\RN{1} \setminus \RN{2}}$, where $\tilde{T}^{+}_{\RN{1} \setminus \RN{2}} = \tilde{T}^{+}_{\RN{1}} \setminus \mathbb{S}^\ast_\RN{2}$ and $\tilde{T}^{-}_{\RN{1} \setminus \RN{2}} = \tilde{T}^{-}_{\RN{1}} \setminus \mathbb{S}^\ast_\RN{2}.$
     
\noindent 3. {\em P-values}:
		For each $t \in \tilde{T}^{+}_{\RN{1} \setminus \RN{2}}$, compute the p-value $p_{\RN{1} \setminus \RN{2}}(t)$ for testing the hypotheses
  		$$
		\begin{aligned}
			\mathcal{H}_{0}(t):& \ \{\mu''(s) = 0 \quad \text{for all} \quad s \in (t-b,t+b) \} \quad {\rm vs.} \\
			\mathcal{H}_{A}(t):& \ \{\mu'(s+) > \mu'(s-) \quad \text{for some} \quad s \in (t-b, t+b) \};
		\end{aligned}
		$$
		and for each $t \in \tilde{T}^{-}_{\RN{1} \setminus \RN{2}}$, compute the p-value $p_{\RN{1}}(t)$ for testing the hypotheses
		$$
		\begin{aligned}
			\mathcal{H}_{0}(t):& \ \{\mu''(s) = 0 \quad \text{for all} \quad s \in (t-b,t+b) \} \quad {\rm vs.} \\
			\mathcal{H}_{A}(t):& \ \{\mu'(s+) < \mu'(s-) \quad \text{for some} \quad s \in (t-b,t+b) \}.
		\end{aligned}
		$$      
        
\noindent 4. {\em Multiple testing}:
		Apply a multiple testing procedure on the set of p-values $\{p_{\RN{1} \setminus \RN{2}}(t), \, t \in \tilde{T}_{\RN{1} \setminus \RN{2}}\}$, and declare significant all local extrema whose p-values are smaller than the significance threshold.
\end{algo}

\subsubsection{Asymptotic FDR control and power consistency}

Let $\mathbb{S}_{1,\RN{1}\setminus\RN{2}} = \cup_{j=1}^{J_1} (v_j - b, v_j +b) \setminus \mathbb{S}^\ast_\RN{2}$ be the signal region of Type \RN{1} change points,
and $\mathbb{S}_{0,\RN{1}\setminus \RN{2}} = U(L) \setminus \mathbb{S}_{1,\RN{1}\setminus \RN{2}}$ be the null region of Type \RN{1} change points.
Then, the number of totally and falsely detected Type \RN{1} change points at threshold $u$ is defined respectively as
$R_{\RN{1} \setminus \RN{2}}(u) = \#\{t\in \tilde{T}_{\RN{1} \setminus \RN{2}}(u)\}$ and  $V_{\RN{1} \setminus \RN{2}}(u) =\#\{t\in \tilde{T}_{\RN{1} \setminus \RN{2}}(u)\cap \mathbb{S}_{0,\RN{1}\setminus \RN{2}}\}$.

Given fixed thresholds $u_1$ and $u_2$ for Type \RN{1} and Type \RN{2} change point detection, respectively, FDR is defined as
$
{\rm FDR}_{\RN{3}}(u_1,u_2) = {\rm E}\left\{\frac{V_{\RN{1} \setminus \RN{2}}(u_1) + V_{\RN{2}}(u_2)}{[R_{\RN{1} \setminus \RN{2}}(u_1) + R_{\RN{2}}(u_2)] \vee1 }\right\},
$
and the power is defined as
\begin{equation} \label{eq:power3}
\begin{split}
{\rm Power}_{\RN{3}}(u_1, u_2) = & \frac{1}{J}\Bigg[\sum_{j\in \mathcal{I}_{\RN{1}}^+} P\left(\tilde{T}_{\RN{1} \setminus \RN{2}}^+(u_1)\cap S_j \ne \emptyset \right) 
 + \sum_{j\in \mathcal{I}_{\RN{1}}^-} P\left(\tilde{T}_{\RN{1} \setminus \RN{2}}^-(u_1)\cap S_j \ne \emptyset \right) \\
 & \quad + \sum_{j\in \mathcal{I}_{\RN{2}}^+} P\left(\tilde{T}_{\RN{2}}^+(u_2)\cap S_j \ne \emptyset \right) 
 + \sum_{j\in \mathcal{I}_{\RN{2}}^-} P\left(\tilde{T}_{\RN{2}}^-(u_2)\cap S_j \ne \emptyset \right)\Bigg] .
\end{split}
\end{equation}
For random thresholds $\tilde{u}_{\RN{1}}$ and $\tilde{u}_{\RN{2}}$ defined in \eqref{eq:thresh-BH-random1} and \eqref{eq:thresh-BH-random2} respectively, ${\rm FDR}_{\RN{3},\rm BH}$ is defined as the FDR with $u_1$ and $u_2$ replaced by $\tilde{u}_{\RN{1}}$ and $\tilde{u}_{\RN{2}}$ respectively.
Similarly, the corresponding power, denoted by ${\rm Power}_{\RN{3},\rm BH}$, is defined as \eqref{eq:power3} with $u_1$ and $u_2$ replaced by $\tilde{u}_{\RN{1}}$ and $\tilde{u}_{\RN{2}}$, respectively.

\begin{theorem}
	\label{thm:type12_fdr}
	Suppose $y(t)$ contains $J_1$ Type \RN{1} change points and $J_2$ Type \RN{2} change points ($J = J_1 + J_2$), conditions (C1) and (C2) hold, and $J_1/L \to A_1>0$ and $J_2/L \to A_2>0 $  as $L\rightarrow \infty$.

\noindent (i)  If Algorithm \ref{alg:type12} is applied with fixed thresholds $u_1$ and $u_2$ for Type \RN{1} and Type \RN{2} change points respectively, then  ${\rm Power}_{\RN{3}}(u_1,u_2) \rightarrow 1$ and $\limsup{\rm FDR}_{\RN{3}}(u_1,u_2)$ are bounded above by
    \begin{equation}
\frac{E[\tilde{m}_{z''_{\gamma}}(U(1),u_1)](1-2c\gamma A_1) + E[\tilde{m}_{z'_{\gamma}}(U(1),u_2)](1-2c\gamma A_2)}
    {E[\tilde{m}_{z''_{\gamma}}(U(1),u_1)](1-2c\gamma A_1) + E[\tilde{m}_{z''_{\gamma}}(U(1),u_2)](1-2c\gamma A_2) + A}.
    \end{equation}
    
\noindent (ii) If Algorithm \ref{alg:type12} is applied with random thresholds $\tilde{u}_{\RN{1}}$ and $\tilde{u}_{\RN{2}}$ for Type \RN{1} and Type \RN{2} change points respectively, then $\limsup{\rm FDR}_{\RN{3},\rm BH} \leq \alpha$ and ${\rm Power}_{\RN{3},\rm{BH}}  \rightarrow 1$.
\end{theorem}

\subsection{Gaussian auto-correlation noise and related peak height distributions}
\label{sec:gau-auto}

Let $X(t)$ be a centered smooth stationary Gaussian process with variance $\sigma^2$. Note that, due to stationarity, $E(X(t)X''(t))=-{\rm Var}(X'(t))$. Define
$
\eta = -{\rm Corr}(X(t), X''(t))=\frac{{\rm Var}(X'(t))}{\sqrt{{\rm Var}(X(t)){{\rm Var}(X''(t))}}}.
$
Let $F_X(x)= P(X(t) > x \ |\  t \ \textnormal{is a local maximum of } X)$ denote the peak height distribution of the Gaussian process $X(t)$. Using the Kac-Rice formula, we obtain 
\begin{align}
F_X(x) = 1 - \Phi\left(\frac{x}{\sigma\sqrt{1-\eta^2}}\right)
+\sqrt{2\pi}\eta \phi\left(\frac{x}{\sigma}\right)\Phi\left(\frac{\eta x}{\sigma\sqrt{1-\eta^2}}\right) \label{eq:peakdist} .
\end{align}
Note that \eqref{eq:peakdist} is a general version of the peak height distribution derived in \citet{schwartzman2011multiple} and \citet{cheng2015distribution}.
An important and attractive characteristic of \eqref{eq:peakdist} is that, for a unit-variance process, it only depends on a single parameter $\eta$, the negative correlation between the process and its second derivative.

To implement the simulations below, we consider a specific example of $X(t)$ and derive the peak height distributions. Let the noise $z(t)$ in \eqref{eq:signal+noise} be $z(t) = \int_{\mathbb{R}} \frac{1}{\nu} \phi\left(\frac{t-s}{\nu}\right)\,dB(s)$ with $\nu > 0$,
where $dB(s)$ is Gaussian white noise ($z(t)$ is regarded by convention as Gaussian white noise when $\nu=0$). Convolving with a Gaussian kernel $w_\gamma(t) = (1/\gamma)\phi(t/\gamma)$ with $\gamma > 0$ produces a zero-mean infinitely differentiable stationary Gaussian process
\begin{equation}\label{eq:gau_stat}
z_\gamma(t) = \int_{\mathbb{R}} w_\gamma(t-x)z(x)dx = \int_{\mathbb{R}} \frac{1}{\xi} \phi\left(\frac{t-s}{\xi}\right)\,dB(s), \quad \xi = \sqrt{\gamma^2 + \nu^2}.
\end{equation}

\begin{lemma}\label{lem:peakheight}
Let $z_{\gamma}(t)$ be defined in \eqref{eq:gau_stat}, the variances of its derivatives are given by
\begin{align*}
    {\rm Var}(z'_{\gamma}(t)) = \frac{1}{4\sqrt{\pi}\xi^3},\ {\rm Var}(z''_{\gamma}(t)) = \frac{3}{8\sqrt{\pi}\xi^5},\ 
    {\rm Var}(z^{(3)}_{\gamma}(t)) = \frac{15}{16\sqrt{\pi}\xi^7}, \ {\rm Var}(z^{(4)}_{\gamma}(t)) = \frac{105}{32\sqrt{\pi}\xi^9}. 
\end{align*}
\end{lemma}

Combining Lemma \ref{lem:peakheight} and \eqref{eq:peakdist}, we immediately obtain the following peak height distributions for the first and second derivatives of the smoothed process.
\begin{proposition}\label{prop:peakdist}
Let $z_{\gamma}(t)$ be defined in \eqref{eq:gau_stat}. Then
, the peak height distributions of $z'_{\gamma}(t) = z^{(1)}_{\gamma}(t)$ and $z''_{\gamma}(t) = z^{(2)}_{\gamma}(t)$ are given respectively by
\begin{equation*}
 F_{z^{(\ell)}_{\gamma}}(x) = 1 - \Phi\left(\frac{x}{\sigma_{\ell}\sqrt{1-\eta_{\ell}^2}}\right)
+\sqrt{2\pi}\eta_\ell \phi\left(\frac{x}{\sigma_\ell}\right)\Phi\left(\frac{\eta_\ell x}{\sigma_\ell\sqrt{1-\eta_\ell^2}}\right), \quad \ell = 1,2,
\end{equation*}
where $\sigma_1^2 = \frac{1}{4\sqrt{\pi} \xi^3}$, $\eta_1 = \frac{\sqrt{3}}{\sqrt{5}}$, $\sigma_2^2 = \frac{3}{8\sqrt{\pi} \xi^5}$ and $\eta_2 = \frac{\sqrt{5}}{\sqrt{7}}$.
\end{proposition}
The p-values in the proposed mSTEM algorithms can be calculated using Proposition \ref{prop:peakdist}. 
Additionally, given that $z_{\gamma}(t)$ is defined in \eqref{eq:gau_stat}, applying the Kac-Rice formula, the expected number of local extrema in the smoothed Gaussian processes $z'_{\gamma}(t)$ and $z^{''}_{\gamma}(t)$ in the unit interval can be calculated by
\begin{align*}
    E[\tilde{m}_{z'_{\gamma}}(U(1))] = \frac{1}{\pi}\sqrt{\frac{{\rm Var}(z_{\gamma}^{(3)}(t))}{{\rm Var}(z_{\gamma}^{''}(t))}} = \frac{\sqrt{10}}{2 \pi \xi }; \quad 
    E[\tilde{m}_{z^{''}_{\gamma}}(U(1))] = \frac{1}{\pi}\sqrt{\frac{{\rm Var}(z_{\gamma}^{(4)}(t))}{{\rm Var}(z_{\gamma}^{(3)}(t))}} = \frac{\sqrt{14}}{2 \pi \xi }.
\end{align*}

\begin{remark}
In this example, Gaussian noise is assumed because it enables a closed formula for the peak height distribution and the expected number of local extrema. For non-Gaussian noise, the p-values associated with the peak heights can be computed via Monte Carlo simulation.
\end{remark}
\subsection{SNR and optimal bandwidth $\gamma$} \label{sec:snr}

Intuitively, a higher SNR facilitates the detection of change points, as it enhances the contrast between signal and noise. 
On the other hand, SNR is not only associated with the size of the slope change and jump, but is also a function of bandwidth $\gamma$. Thus, it is important to study the SNR when choosing the best bandwidth. The following result, which is a direct consequence of Lemmas \ref{lemma:u1} and \ref{lem:peakheight}, provides definitions and properties for the SNR of both types of change points.


\begin{lemma}\label{lem:snr}
Suppose that $z_{\gamma}(t)$ is the smoothed noise defined in \eqref{eq:gau_stat}, we let $\nu = m \gamma$ ($m=0$ when $z(t)$ is Gaussian white noise).
For a Type \RN{1} change point $v_j$, the {\rm SNR} is calculated as 
\begin{align*}
\text{\rm SNR}_{\RN{1}}(v_j) := \frac{|\mu''_{\gamma}(v_j)|}{\sqrt{{\rm Var}(z''_{\gamma}(v_j))}} = \frac{|\frac{k_{j+1}-k_j}{\sqrt{2\pi}\gamma}|}{\sqrt{\frac{3}{8\sqrt{\pi}\xi^5}}}
= \frac{2\gamma^{3/2}}{\sqrt{3}\pi^{1/4}}(1+m^2)^{5/4} |k_{j+1}-k_j|.
\end{align*}
For a Type \RN{2} change point $v_j$, it asymptotically becomes a local extremum at $v_j$ under condition (C2), and the corresponding {\rm SNR} is calculated as
\begin{align*}
&\text{\rm SNR}_{\RN{2}}(v_j) := \frac{|\mu'_{\gamma}(v_j)-k_j|}{\sqrt{{\rm Var}(z'_{\gamma}(v_j))}} = \frac{|\frac{a_j}{\sqrt{2\pi}\gamma}+\frac{k_{j+1}-k_j}{2} +(k_j+k_{j+1})(\Phi(c)-1)|}{\sqrt{\frac{1}{4\sqrt{\pi}\xi^3}}}\\
&= |\frac{\sqrt{2}a_j}{\pi^{1/4}}(1+m^2)^{3/4}\sqrt{\gamma} + [(k_{j+1}-k_j)+2(k_j+k_{j+1})(\Phi(c)-1)]\pi^{1/4}(1+m^2)^{3/4}{\gamma}^{3/2} |.
\end{align*}
\end{lemma}

Due to condition (C2) and $\Phi(c)\approx 1$, the SNR for both Type \RN{1} and \RN{2} change points is increasing in the sizes of slope change and jump, respectively. 
In addition, Lemma \ref{lem:snr} shows that the SNR increases in $\gamma$ for both Type \RN{1} and Type \RN{2} change points. Therefore, if the minimal distance between the neighboring change points $d$ is large enough, a large $\gamma$ is preferable (see Figure \ref{fig:type12}). However, for a fixed $d$, to avoid the overlap of two neighboring smoothed signal segments $(v_j -c\gamma, v_j + c\gamma)$ and $(v_{j+1}-c\gamma, v_{j+1}+c\gamma)$, $\gamma$ should not be too large (such as $2c\gamma \leq d$). 


\section{Simulation Study}
\label{sec:simulation}

\subsection{Simulation settings}\label{sec:simu_set}

In this section, we will assess the performance of our method in various signal scenarios. The signals follow the form $\mu(t) = c_j + k_j t$,
where $t = 1,\dots,L$. True change points are defined as $v_j = jd$ for $j=1,\dots,\lfloor L/d \rfloor -1 $ with $d=150$ representing the separation between adjacent change points.
We consider four distinct signal scenarios with corresponding signal coefficients:
(1). piecewise linear mean with continuous change points (Type \RN{1} change points): $c_1=k_1=0$ and $k_{j+1} - k_j = 0.1$ for $j\geq 1$;
(2). piecewise constant mean with jumps (special case of Type \RN{2} change points): $c_1=k_1=0$ and $a_j = c_{j+1}-c_j = 10$ for $j\geq 1$;
(3). piecewise linear mean with discontinuous change points (Type \RN{2} change points): $c_1=k_1=0$, $a_j=10$, $k_{j+1}-k_j=0.05$ for odd $j$ and $k_{j+1}-k_j=-0.05$ for even $j$;
(4). mixture of Type \RN{1} and Type \RN{2} change points: This scenario consists of the signal from scenario (1) followed by the signal from scenario (3).
For the first three scenarios, let $L = 1500$ and thus $J = 9$ for each scenario. For the mixture case, we extend the signal length to $L = 3000$ with $J_1 = J_2 = 9$.

The noise is generated from a stationary ergodic Gaussian process with zero mean and  $\nu=1$.
To control the False Discovery Rate (FDR), we implement the BH procedure at a significant level of $\alpha = 0.05$.
Our simulation results are based on the average of 1000 replications.

\subsection{Performance of our method}

\begin{figure}[!ht]
	\centering
	\includegraphics[width=0.70\textwidth]{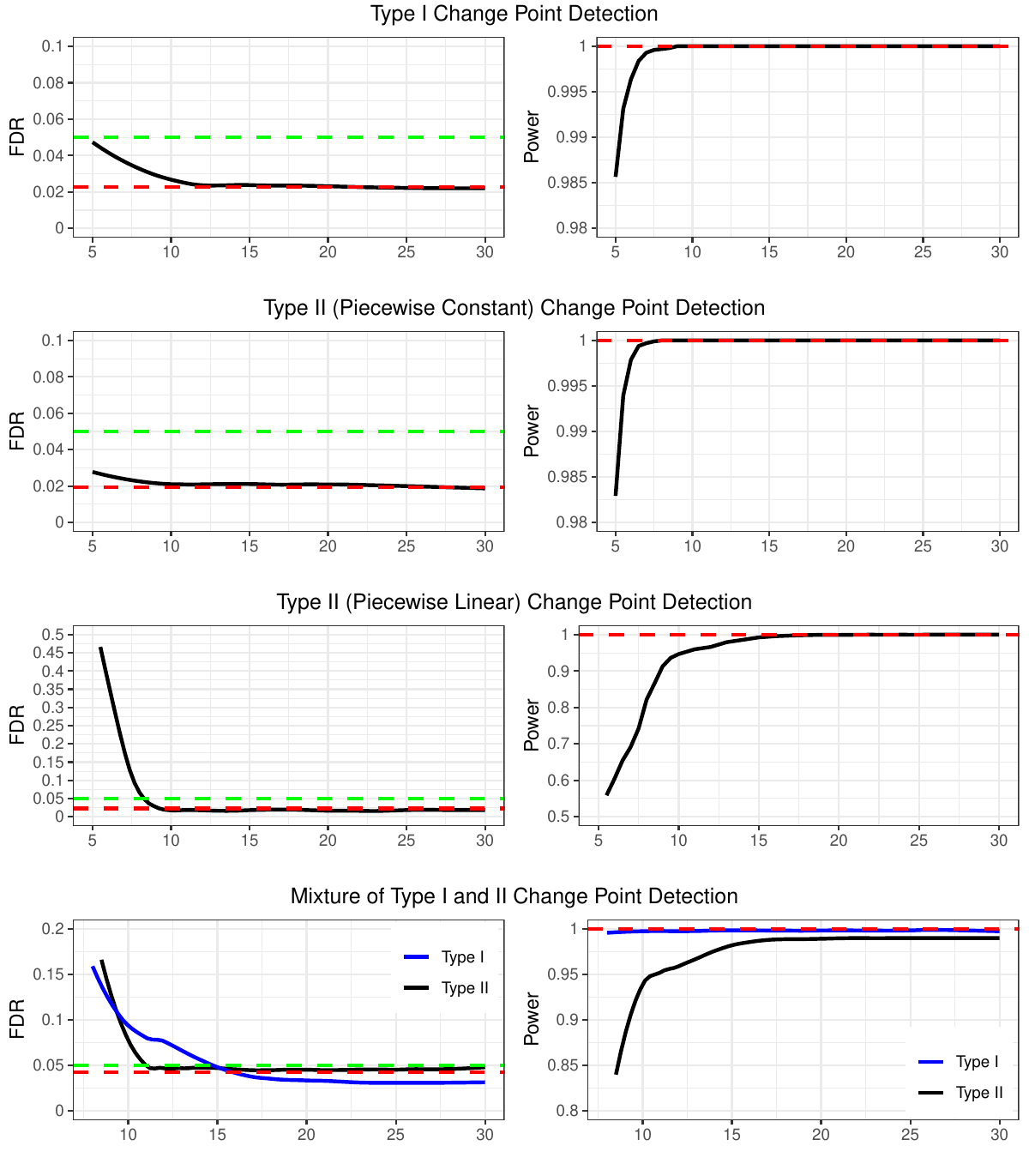} 
	\caption{\footnotesize FDR (left panel) and power (right panel) versus SNR for detecting change points in various signal types. 
	This example employs a kernel bandwidth of $\gamma=10$ and a location tolerance of $b=10$.
	The red dashed lines mark the asymptotic limits of FDR and power, while the green dashed lines indicate the significant level $\alpha=0.05$.}
	\label{fig:type12}
\end{figure}

In Section \ref{sec:multiple_test}, we presented the theoretical results of the False Discovery Rate (FDR) and the power of our method. 
In this section, we validate these theoretical properties through numerical studies.

Figure \ref{fig:type12} shows the performance of our method in four different signal scenarios. 
Notably, as SNR increases, the FDR converges to its asymptotic limit. This limit remains consistently below the FDR control level of $\alpha=0.05$. Meanwhile, the statistical power approaches 1.
For a fixed bandwidth $\gamma$, a higher SNR implies a more substantial slope change or jump size, aligning with conditions (C1) and (C2), respectively.
It is worth noting that while our theoretical analysis assumes an infinite SNR, our numerical results show that even with a moderate SNR, both the FDR and power have already converged to their asymptotic limits.

\subsection{Comparison with other methods}

To compare with other change point detection methods, we evaluated the precision and computational efficiency of change point detection. Short-term (see section \ref{sec:simu_set}) and long-term data sequences were considered.
The extended long-term data sequence was generated by replicating the short-term data tenfold within our simulation, guaranteeing that the underlying patterns remained consistent between the two datasets. 

We compared our method with three multiple testing-based change point detection methods: B\&P \citep{bai1997estimation}, NOT \citep{baranowski2019narrowest}, and NSP \citep{fryzlewicz2020narrowest}. 
Note that we excluded the results of the B\&P method from the long-term data comparison due to the extremely long computing time. For the NSP method, the change points were estimated as confidence intervals, and we used the midpoints of these intervals as point estimates for comparison with other methods for the sake of consistency.
Furthermore, the scenario of a mixture of change points of Type \RN{1} and Type \RN{2} was not considered because the other three methods cannot distinguish between change points of Type \RN{1} and Type \RN{2}.

Suppose that the number of true change points is $J$, and the number of detected change points is $\hat{J}$.
For each detected change point $\hat{v}_i$, $i = 1,2,\dots,\hat{J}$, we define it as the estimate of the true change point $v_j$ such that $j = \argmin_{j=1,2,\dots,J} |\hat{v}_i - v_j|$.
Thus, we let $v_i^0 = v_j$ be the true change point corresponding to $\hat{v}_i$. Note that a single true change point may correspond to multiple estimated change points.
The detection accuracy of $\hat{v}_i$ is measured based on the distance $|\hat{v}_i - v_i^0|$ for $i=1,2,\dots,\hat{J}$,
and the capture rate of an interval $I$ is defined as $\sum_{i=1}^{\hat{J}} \mathbbm{1}(|\hat{v}_i - v_i^0| \in I)/J$.
A high capture rate may be achieved by detecting an excessive number of change points, which, in turn, leads to a high FDR. Therefore, a good change point detection method should achieve high detection accuracy while maintaining a low FDR.

In Table \ref{tab:compare_short}, the results for the short-term data illustrate the performance of our proposed method, NOT and NSP in terms of FDR control and power in all scenarios. Both NOT and our method demonstrate the capability to achieve high accuracy in change point detection within a single bandwidth $\gamma$, and our method is the fastest among the algorithms considered.
In Table \ref{tab:compare_long}, it is seen that for the long-term data sequence, NOT and NSP do not perform well with respect to FDR control, power, and detection accuracy. In contrast, our method consistently maintains excellent performance, retaining the advantage of the shortest computing time.
Note that the better performance of our proposed method in terms of FDR and power with long-term data can be attributed to asymptotic conditions $L\rightarrow \infty$ and $J/L \rightarrow A > 0$.
Additionally, the minimal computing time required by our method in both short-term and long-term data sequences demonstrates its superior computational complexity.

\begin{table}[!ht]
\centering
\footnotesize
\caption{\footnotesize Detection accuracy of change points for the short-term data sequence ($\gamma = b = 10$).
}
\label{tab:compare_short}
\begin{threeparttable}
\begin{tabular}{cccccccccc}
\toprule
\multirow{2}*{Signal Type} & \multirow{2}*{Method} & \multicolumn{5}{c}{Capture rate of interval} & \multirow{2}*{FDR} & \multirow{2}*{Power} & \multirow{2}*{Time (s)}\\
  \cline{3-7}
  ~ & ~ & $[0,\frac{1}{3}\gamma)$ & $[\frac{1}{3}\gamma,\gamma)$ & $[\gamma,2\gamma)$ &
  $[2\gamma,4\gamma)$ & $\geq 4\gamma$ & ~ \\
\midrule
\multirow{4}*{Type \RN{1}} & mSTEM & 0.8400 & 0.1533 & 0.0217 & 0.0267 & 0.0483 &0.0125 &0.9933 & 0.1370\\
 & NOT & 0.9883 & 0.0117 & 0.0000 & 0.0017 & 0.0000 &0.0572 &1.0000 & 1.3550\\
 & NSP & 0.6183 & 0.1900 & 0.1617 & 0.0300 & 0.0000 &0.0458 &0.9999 & 3.2417\\
 & B\&P & 0.4700 & 0.5083 & 0.0217 & 0.0000 & 0.0000 &0.1632 &0.9783 & 87.4562\\
\midrule
\multirow{4}*{\makecell[c]{Type \RN{2}\\Piecewise Constant}} & mSTEM & 0.9617 & 0.0383 & 0.0000 & 0.0367 & 0.0333 &0.0227 &1.0000 & 0.0290\\
 & NOT & 0.9833 & 0.0183 & 0.0017 & 0.0000 & 0.0017 &0.0558 &1.0000 & 0.2863\\
 & NSP & 0.6767 & 0.3133 & 0.2117 & 0.0867 & 0.0100 &0.0517 &0.9001 & 1.9430\\
 & B\&P & 0.4117 & 0.0233 & 0.0467 & 0.2050 & 0.0000 &0.1260 &0.4350 & 71.0924\\
\midrule
\multirow{4}*{\makecell[c]{Type \RN{2}\\Piecewise Linear}} & mSTEM & 0.9983 & 0.0017 & 0.0000 & 0.0067 & 0.0233 &0.0348 &1.0000 & 0.0839\\
 & NOT & 0.8833 & 0.0933 & 0.0233 & 0.0000 & 0.0000 &0.0727 &0.9766 & 0.4524\\
 & NSP & 0.6967 & 0.2417 & 0.0333 & 0.0283 & 0.0000 &0.0626 &0.9384 & 2.8013\\
 & B\&P & 0.3333 & 0.0000 & 0.3015 & 0.3333 & 0.0745 &0.1633 &0.3333 & 68.3345\\
\midrule
\multirow{4}*{\makecell[c]{Type \RN{2}\\Piecewise Linear}} & mSTEM & 0.9983 & 0.0017 & 0.0000 & 0.0067 & 0.0233 &0.0348 &1.0000 & 0.0839\\
& NOT & 0.8833 & 0.0933 & 0.0233 & 0.0000 & 0.0000 &0.0727 &0.9766 & 0.4524\\
& NSP & 0.6967 & 0.2417 & 0.0333 & 0.0283 & 0.0000 &0.0626 &0.9384 & 2.8013\\
\bottomrule
\end{tabular}
\end{threeparttable}
\end{table}

\begin{table}[!ht]
\centering
\footnotesize
\caption{ \footnotesize Detection accuracy of change points for the long-term data sequence ($\gamma = b = 10$).}
\label{tab:compare_long}
\begin{threeparttable}
\begin{tabular}{cccccccccc}
\toprule
\multirow{2}*{Signal Type} & \multirow{2}*{Method\tnote{\textdagger}} & \multicolumn{5}{c}{Capture rate of interval}  & \multirow{2}*{FDR} & \multirow{2}*{Power} & \multirow{2}*{Time (s)}\\
  \cline{3-7}
  ~ & ~ & $[0,\frac{1}{3}\gamma)$ & $[\frac{1}{3}\gamma,\gamma)$ & $[\gamma,2\gamma)$ &
  $[2\gamma,4\gamma)$ & $\geq 4\gamma$ & ~ & ~ & ~\\
\midrule
\multirow{3}*{Type \RN{1}} & mSTEM & 0.7616 & 0.2241 & 0.0295 & 0.0244 & 0.0135 & 0.0127 & 0.9963 & 0.2469\\
 & NOT & 0.1358 & 0.1712 & 0.2475 & 0.5028 & 0.2369 &0.0853 &0.8732 & 112.5007\\
 & NSP & 0.3512 & 0.4607 & 0.1692 & 0.0086 & 0.0000 &0.0792 &0.8362 & 433.6193\\
\midrule
\multirow{3}*{\makecell[c]{Type \RN{2}\\Piecewise Constant}} & mSTEM & 0.9702 & 0.0000 & 0.000 & 0.0154 & 0.0161 &0.01463 &1.0000 & 0.1188\\
 & NOT & 0.8633 & 0.0037 & 0.007 & 0.0177 & 0.0090 &0.0735 &0.9164 & 48.4487\\
 & NSP & 0.6398 & 0.3202 & 0.030 & 0.0000 & 0.0000 &0.08011 &0.9228 & 213.0398\\
\midrule
\multirow{3}*{\makecell[c]{Type \RN{2}\\Piecewise Linear}} & mSTEM & 0.9899 & 0.0000 & 0.0000 & 0.0065 & 0.0133 &0.0237 &0.9992 & 1.0627\\
 & NOT & 0.8748 & 0.0002 & 0.0009 & 0.0032 & 0.0025 &0.1217 &0.8012 & 10.6064\\
 & NSP & 0.6059 & 0.3543 & 0.0298 & 0.0000 & 0.0000 &0.1383 &0.8255 & 400.6633\\
\bottomrule
\end{tabular}
\begin{tablenotes}
\item [\textdagger] {\footnotesize B\&P method was not included due to its substantial computational time requirements}.
\end{tablenotes}
\end{threeparttable}
\end{table}

\section{Data examples}
\label{sec:data}

In this section, we applied our method to real applications and performed a comparative analysis with NOT and NSP.
We evaluated the performance of all three methods on the dataset of global temperature records.

\subsection{Global temperature records}

Climate change research plays a pivotal role in addressing the impacts of global warming and guiding policy decisions related to mitigation and adaptation strategies. 
Analyzing temperature changes is crucial for understanding the patterns and shifts in climate over time.
In this paper, we study the data concerning global mean land-ocean temperature deviations (measured in degrees Celsius) from the period 1880 to 2015, in comparison to the average temperature between 1951 and 1980. This dataset is available at \url{https://data.giss.nasa.gov/gistemp/graphs}.
\begin{figure}[!ht]
    \centering
    \includegraphics[width=0.50\textwidth]{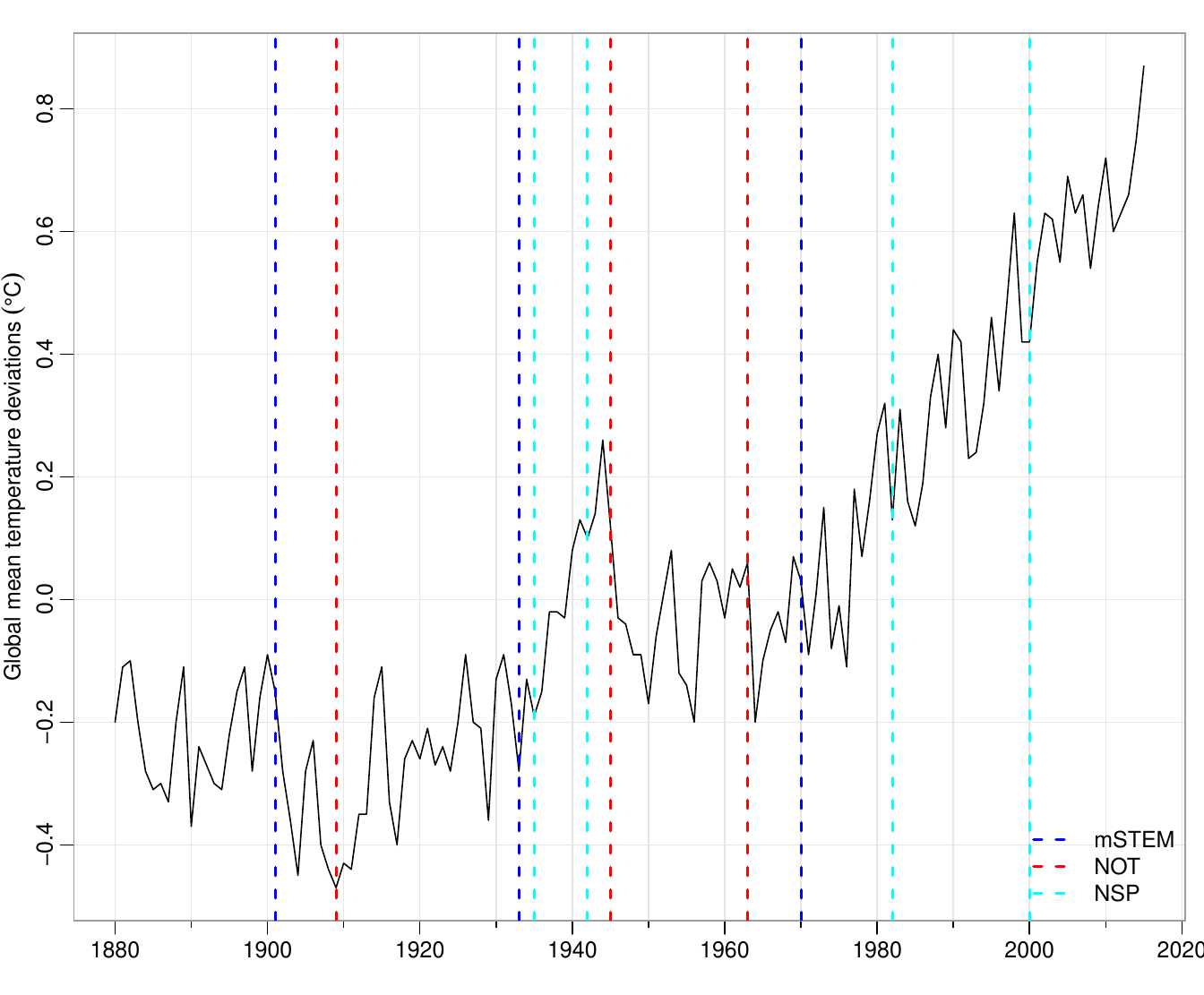}
    \caption{\footnotesize Change point detection of global mean land-ocean temperature deviations.}
    \label{fig:globtemp}	
\end{figure}

Figure \ref{fig:globtemp} illustrates the results of applying all three methods to global temperature data. Our method successfully detected two Type \RN{2} change points in 1902 and 1934, along with one Type \RN{1} change point in 1971. 
These findings closely align with the results presented in a previous study by \citet{yu2019change}, where they conducted an analysis of five time series and identified three distinct periods of significant temperature change in the 20th century. Specifically, their research highlighted change points between 1902-1917, 1936-1945, and 1963-1976.  
The result of our detection method aligns with the visual assessment of the data sequence.
The observed pattern suggests that the signal remained relatively constant during the first period (1880 - 1902), followed by a linear trend in the second period (1903 - 1934), then another constant signal in the third period (1935 - 1971), and finally a linear trend in the last period.

\section{Discussion}
\label{sec:discussion}

\subsection{Choice of smoothing bandwidth $\gamma$}

The choice of the smoothing bandwidth $\gamma$ is, in fact, an important aspect of change point detection. It plays a role in balancing the trade-off between the overlap of the smoothed signal supports and the accuracy of detecting change points. 
Thus, a small $\gamma$ (if the noise is highly autocorrelated) or a relatively large $\gamma$ (if the noise is less autocorrelated) is preferred to increase the power, but only to the extent that the smoothed signal supports have little overlap. 
Considering the Gaussian kernel with support domain $\pm c\gamma$, an initial guideline to choose $\gamma$ can be set as $d/(2c)$, where d is the minimum distance between any two neighboring change points. This guideline ensures that the supports of the smoothed signals do not overlap excessively. In practice,  a data-driven approximation of $d$  is to use an initial estimator from some classic change point detection methods, such as NOT or NSP, and the resulting $\widehat{d}$ provides a practical anchor for bandwidth selection. A reasonable choice is then $\gamma \approx \widehat{d}/(2c)$, which helps prevent excessive overlap of smoothed signal supports.



\subsection{Type \RN{2} change points with moderate $|q_j|$}

For a Type \RN{2} change point $v_j$, we have studied in this paper the case of small $|q_j|$, and we showed in Section S1 of the Supplementary Materials the case of large $|q_j|$, which behaves similarly to a Type \RN{1} change point. But when a Type \RN{2} change point has a moderate $|q_j|$, for example, $|q_j|$ is close to $c/\gamma$, the first derivative $\mu'_{\gamma}(t)$ can produce a local extremum at $t = v_j \pm \gamma^2 |q_j| \approx v_j \pm c\gamma$, which is near the endpoints of the smoothed signal region and outside of $v_j \pm b$.
This may lead to false detections in the noise region while failing to detect 
$v_j$ in the signal region, resulting in an overestimation of FDR and an underestimation of power.
Addressing this limitation of detecting Type \RN{2} change points with moderate $|q_j|$ would be an interesting question for future research.

\subsection{Nonstationary Gaussian noise}

In this paper, the Gaussian noise is assumed to be stationary, which allows the error terms to be correlated. However, in real applications, it is common for the data to contain nonstationary Gaussian noise \citep{heinonen2016non}. Thus, it is valuable to detect change points under nonstationary Gaussian noise. To achieve this, we will study the peak height distribution for nonstationary Gaussian noise and show the FDR control and power consistency under certain assumptions, which would be promising but comes with specific challenges.

\section*{Acknowledgments}
The authors thank Prof. Domenico Marinucci for motivating this research problem and Prof. Armin Schwartzman for helpful discussions and valuable suggestions.
 The authors also thank the Editor, Associate Editor and the reviewers for their efforts and suggestions.

\section*{Funding}
Dan Cheng acknowledges support from the National Science Foundation grant DMS-2220523 and the Simons Foundation Collaboration Grant 854127.

\bibliography{chp} 
\bibliographystyle{agsm}

\end{document}